\documentclass[twocolumn]{aastex61}
\usepackage{amsmath}
\begin{document}

\newcommand{\Mbh}{$M_{\rm BH}$}
\newcommand{\g}{\textit{g}}
\newcommand{\iband}{\textit{i}-bands}
\accepted{for publication in APJ}

\title{The Sloan Digital Sky Survey Reverberation Mapping Project: Accretion-Disk Sizes from Continuum Lags}
\correspondingauthor{Yasaman Homayouni}
\email{yasaman.homayouni@uconn.edu}

\author{Y. Homayouni}
\affil{University of Connecticut, Department of Physics, 2152 Hillside Road, Unit 3046, Storrs, CT 06269-3046}

\author{Jonathan R. Trump}
\affiliation{University of Connecticut, Department of Physics, 2152 Hillside Road, Unit 3046, Storrs, CT 06269-3046}

\author{C. J. Grier}
\affiliation{Dept. of Astronomy and Astrophysics, The Pennsylvania State University, 525 Davey Laboratory, University Park, PA 16802}
\affiliation{Institute for Gravitation and the Cosmos, The Pennsylvania State University, University Park, PA 16802}
\affiliation{Steward Observatory, The University of Arizona, 933 North Cherry Avenue, Tucson, AZ 85721, USA}

\author{Yue Shen}
\affiliation{Department of Astronomy, University of Illinois at Urbana-Champaign, Urbana, IL, 61801, USA}
\affiliation{National Center for Supercomputing Applications, University of Illinois at Urbana-Champaign, Urbana, IL, 61801, USA}

\author{D. A. Starkey}
\affiliation{Department of Astronomy, University of Illinois at Urbana-Champaign, Urbana, IL, 61801, USA}

\author{W. N. Brandt}
\affiliation{Dept. of Astronomy and Astrophysics, The Pennsylvania State University, 525 Davey Laboratory, University Park, PA 16802}
\affiliation{Institute for Gravitation and the Cosmos, The Pennsylvania State University, University Park, PA 16802}
\affiliation{Department of Physics, 104 Davey Lab, The Pennsylvania State University, University Park, PA 16802, USA}

\author{G. Fonseca Alvarez}
\affiliation{University of Connecticut, Department of Physics, 2152 Hillside Road, Unit 3046, Storrs, CT 06269-3046}

\author{P. B. Hall}
\affiliation{Department of Physics and Astronomy, York University, Toronto, ON M3J 1P3, Canada}

\author{Keith Horne}
\affiliation{SUPA Physics and Astronomy, University of St. Andrews, Fife, KY16 9SS, Scotland, UK}

\author{Karen Kinemuchi}
\affiliation{Apache Point Observatory and New Mexico State University, P.O. Box 59, Sunspot, NM, 88349-0059, USA}

\author{Jennifer I-Hsiu Li}
\affiliation{Department of Astronomy, University of Illinois at Urbana-Champaign, Urbana, IL, 61801, USA}

\author{Ian D. McGreer}
\affiliation{Steward Observatory, The University of Arizona, 933 North Cherry Avenue, Tucson, AZ 85721, USA}

\author{Mouyuan Sun}
\affiliation{CAS Key Laboratory for Research in Galaxies and Cosmology, Department of Astronomy, University of Science and Technology of China, Hefei 230026, China}
\affiliation{School of Astronomy and Space Science, University of Science and Technology of China, Hefei 230026, China}

\author{L.C. Ho}
\affiliation{Kavli Institute for Astronomy and Astrophysics, Peking University, Beijing 100871, China}
\affiliation{Department of Astronomy, School of Physics, Peking University, Beijing 100871, China}

\author{D. P. Schneider}
\affiliation{Dept. of Astronomy and Astrophysics, The Pennsylvania State University, 525 Davey Laboratory, University Park, PA 16802}
\affiliation{Institute for Gravitation and the Cosmos, The Pennsylvania State University, University Park, PA 16802}

\begin{abstract}
We present accretion-disk structure measurements from continuum lags in the Sloan Digital Sky Survey Reverberation Mapping (SDSS-RM) project. Lags are measured using the \texttt{JAVELIN} software from the first-year SDSS-RM $g$ and $i$ photometry, resulting in well-defined lags for 95 quasars, 33 of which have lag SNR $>$ 2$\sigma$. We also estimate lags using the \texttt{ICCF} software and find consistent results, though with larger uncertainties. Accretion-disk structure is fit using a Markov Chain Monte Carlo approach, parameterizing the measured continuum lags as a function of disk size normalization, wavelength, black hole mass, and luminosity. In contrast with previous observations, our best-fit disk sizes and color profiles are consistent (within 1.5~$\sigma$) with the \citet{SS73} analytic solution. We also find that more massive quasars have larger accretion disks, similarly consistent with the analytic accretion-disk model. The data are inconclusive on a correlation between disk size and continuum luminosity, with results that are consistent with both no correlation and with the \citet{SS73} expectation. The continuum lag fits have a large excess dispersion, indicating that our measured lag errors are underestimated and/or our best-fit model may be missing the effects of orientation, spin, and/or radiative efficiency. We demonstrate that fitting disk parameters using only the highest-SNR lag measurements biases best-fit disk sizes to be larger than the disk sizes recovered using a Bayesian approach on the full sample of well-defined lags.
\end{abstract}


\section{Introduction} \label{sec:intro}
Quasars are supermassive black holes (SMBHs) that grow by rapid mass accretion. During the accretion phase quasars glow with total luminosity $L_{\rm Bol} = \eta \dot{M}c^2$, where $\eta$ is the radiative efficiency, $\dot{M} = dM/dt$ is the SMBH accretion rate, and $c$ is the speed of light. The foundational model for black hole accretion disks is the thin-disk model of \citet[][hereafter SS73]{SS73}. The SS73 disk model is an optically thick, geometrically thin disk model where the local disk emission corresponds to a series of black bodies at different radii. The inner part of the accretion disk has hotter emission whereas at the outer edge of the disk the emission is cooler.

Even though the SS73 model has been widely used, mounting observational and theoretical evidence shows that the SS73 disk model breaks down in several ways. Recent continuum reverberation mapping (RM) observations \citep{Shappee2014,Fausnaugh2016,Fausnaugh2017,Jiang2017a, Mudd2017} identified discrepancies in the measured disk sizes from what is expected by the SS73 model. This discrepancy is also reported in micro-lensing observations of quasars \citep{Morgan2010,Blackburne2011,Motta2017}. 

Both theory and non-RM observations suggest that black hole accretion structure depends on accretion rate in ways that are not entirely predicted by the SS73 thin-disk model. Recent advances in simulations of super-Eddington accretion disks predict dramatically different emission and outflow properties compared to the sub-Eddington SS73 analytic prescription (\citealt{Sadowski2014, Sadowski2016, McKinney2014, Jiang2014, Jiang2017b}, see also the analytic ``slim" disk model of \citealt{Abramowicz1988}). Observations of candidate super-Eddington quasars in X-ray \citep{Desroches2009}, with broad-line kinematics \citep{Du2015} and spectral energy distribution (SED) fitting \citep{Luo2015} show similar evidence for slim accretions disks.
At low accretion rates, SED observations suggest that accretion occurs in a hot, ionized, optically thin, radiatively inefficient accretion flow (RIAF) mode, although the exact radiative efficiency is degenerate with the mass accretion rate \citep{Narayan1994,Narayan2008, Ho2008, Trump2011, Elitzur2014}. Additional theoretical work suggests that different wind profiles can cause the disk structure and emission properties to differ from the SS73 model \citep{Slone2012, Laor2014, Sun2018c}.

Testing the connections between accretion-disk size, $M_{\rm BH}$ and $\dot{M}$ may reveal whether the ratio of observational to theoretical disk sizes depends on $M_{\rm BH}$ and / or accretion rate. These ideas have not yet been tested by direct accretion-disk measurements, since previous reverberation mapping surveys provide measurements for only small samples spanning a narrow range of black hole mass and accretion rate estimates. The SS73 thin blackbody disk model predicts that the disk size, $r = c\,\tau$, at rest-frame wavelength $\lambda$ depends on the black hole mass $M_{\rm BH}$ and accretion rate $\dot{M}$, both with a power-law index of 1/3, as follows:
\begin{equation}\label{eq:SS73_theory}
c\tau = \Big(\frac{45\,G}{16\,\pi^6\,h\,c^2}\Big)^{1/3}\,\lambda^{4/3}\, M_{\rm BH}^{1/3}\, \dot{M}^{1/3}
\end{equation}

The bulk of underlying accretion physical processes occurs within light-years of the central black hole, which cannot be spaitially resolved with current technology. The RM method \citep{Blandford1982, Peterson2004} is a powerful tool for investigating regions where direct imaging cannot resolve structure. The RM method substitutes high temporal resolution for high spatial resolution, allowing us to probe regions that are only ∼light-days in extent. RM is enabled by the fact that quasar luminosity is variable, and we observe physically connected regions ``reverberate" in response to the driving continuum. The variability signatures in high-energy emission regions are thus repeated in lower-energy emission regions, with the signals delayed by the time required for the light to travel between the two regions\footnote{As is standard in reverberation mapping studies, we assume a ``lamp post" model where fluctuations are driven at the speed of light \citep{Cackett2007}.  Other mechanisms for driving fluctuations with $v \ll c$, like sound waves, would imply implausibly small disks.  We also assume that the distance between wavelength regions remains constant during luminosity fluctuations, consistent with the relatively small (average $\sim$8\%) rms variability of the continuum light curves.}. The RM technique is most frequently applied to measure the time delay between variations in the observed-frame optical continuum emission and the broad emission lines emitted in the eponymous broad-line region. This time delay yields the relative sizes of each of these regions. Broad-line RM is currently the only method to robustly measure SMBH mass in active galaxies beyond $\sim$ 100 Mpc.

Continuum RM \citep{Krolik1991, Fausnaugh2016} measures the variability of the continuum emission at various wavelengths in response to the driving UV/X-ray ionizing continuum. Measuring the variability in the re-emitted continuum emission from the accretion disk probe the accretion disk regions that emit black body radiation. Continuum lags at different wavelengths, resulting from the emission of hotter regions closer to the black hole, and cooler more distant disk regions, can be used to measure disk sizes. In addition, by measuring the response of the continuum emission from different parts of the disk, one can map the temperature and wavelength scaling of the accretion-disk structure.

Previous continuum RM campaigns have dedicated many observations to interband optical monitoring \citep{Sergeev2005, Cackett2007} and a few have even been extended to UV and soft/hard X-ray \citep{Wanders1997,Collier1998, Gehrels2004, Shappee2014, McHardy2014, Fausnaugh2016,Edelson2017, McHardy2018}. These previous results, based on cross-correlation lag measurements, are consistent with the $T\propto r^{-3/4}$ and thus $\tau \propto \lambda^{4/3}$ prediction of the SS73 model (although see also \citep{Starkey2017}). Continuum RM observations also find a measured disk normalization that is $\approx$ 3-4 times larger than expected \citep{Edelson2015, Edelson2017, Jiang2017a, Fausnaugh2016}. This result is also in agreement with microlensing observations \citep{Morgan2010, Blackburne2011, Motta2017}. Recently, \citet{Mudd2017} report lag upper limits from the Dark Energy Survey consistent with the SS73 model assuming moderate to high accretion rates.

The inhomogeneous disk models explained by \citet{Dexter2011} incorporate temperature fluctuations in Keplerian rotation disks that can produce larger disk sizes; in addition this would solve the problem of quasar variability that is not well understood in the context of the SS73 model. However, previous studies have not tested disk-structure dependency on $M_{\rm BH}$ and accretion rate due to current data limited to low-luminosity Seyfert galaxies. There are currently only seven Type 1 Seyfert AGNs that have both continuum and emission-line RM measurements, which together allow for both direct $M_{\rm BH}$ and accretion-disk size measurements \citep{Collier1998, Edelson2015, Fausnaugh2016, Edelson2017, McHardy2018, Fausnaugh2018}.

We address this problem by performing a comprehensive study of the physics of black hole accretion using direct accretion-disk size and structure measurements from the Sloan Digital Sky Survey Reverberation Mapping (SDSS-RM) project \citep{Shen2015} between optical \textit{g} and \textit{i} photometry bands. We connect the observed accretion-disk structure with black hole mass and accretion rate using our unique sample of quasars that have well-measured BH masses from a previous SDSS-RM BH mass study \citep{Grier2017}. This work is complementary to Starkey et al. (in prep), which uses a different methodology to similar measure continuum lags from SDSS-RM quasars.  Here we focus on measuring disk size, color profile, and disk dependence on mass and luminosity using the \texttt{JAVELIN} software, which fits reverberation lags using damped random walk (DRW) model for the statistical behavior of lightcurve variability. In contrast, Starkey et al. (in prep) uses the \texttt{CREAM} software with models for both the driving lightcurve and the disk reverberation response, fitting disk size, temperature profile, and orientation. Section 2 describes our sample chosen from the SDSS-RM dataset. Section 3 presents our procedure for lag identification, including alias removal, outlier rejection and lag quality analysis. In section 4 we discuss the necessary criteria for selecting physical lags corresponding to reverberating light curves. Section 5 describes our use of computed lags to fit a normalization of the accretion disk and link the observed lags to mass and accretion rate correlations. Throughout this work, we adopt a $\Lambda$CDM cosmology with $\Omega_{\Lambda}$ = 0.7, $\Omega_{M}$ = 0.3, and h = 0.7.
\section{Data}\label{sec:data}
\subsection{SDSS-RM Survey}
The Sloan Digital Sky Survey Reverberation Mapping project (SDSS-RM) is a pioneering 
multi-object RM campaign \citep{Shen2015} that is simultaneously monitoring a sample of 849 quasars in a single 7 deg$^2$ field since 2014, the project began with SDSS-III \citep{Eisenstein2011}. The selected RM sample is flux-limited to $i_{psf} = 21.7$ with no additional cuts on variability amplitude or redshift of the quasars, dramatically expanding the parameter space of spectroscopic, variability and multi-wavelength properties of quasars with RM data (Figure 1 of Shen et al. 2015). The main goal of SDSS-RM is to measure lags for a range of emission lines and measure black hole mass, as well as improving the established radius-luminosity (R-L) relation \citep{Kaspi2007, Bentz2013} that is currently well-calibrated for H$\beta$ in a biased sample of nearby $z<0.3$ quasars. Due to the necessity of continuous observations in this survey, coordinated monitoring by different SDSS-RM photometry sites is essential to monitor quasar light variability. Thus the SDSS-RM program is supported by ground-based photometry from multiple facilities including the Canada-France-Hawaii Telescope (CFHT) and Steward Observatory Bok telescope. 
To date, SDSS-RM has resulted in several studies of the variability and properties of quasar emission lines \citep{Sun2015,Denney2016a,Shen2016b,Denney2016b,Li2017,Sun2018}, broad absorption line variability \citep{Grier2016}, the relationship between black hole growth and host galaxy properties and broad emission-line lags \citep{Matsuoka2015,Shen2016a,Grier2017}.

\begin{figure}
\includegraphics[scale = 0.30]{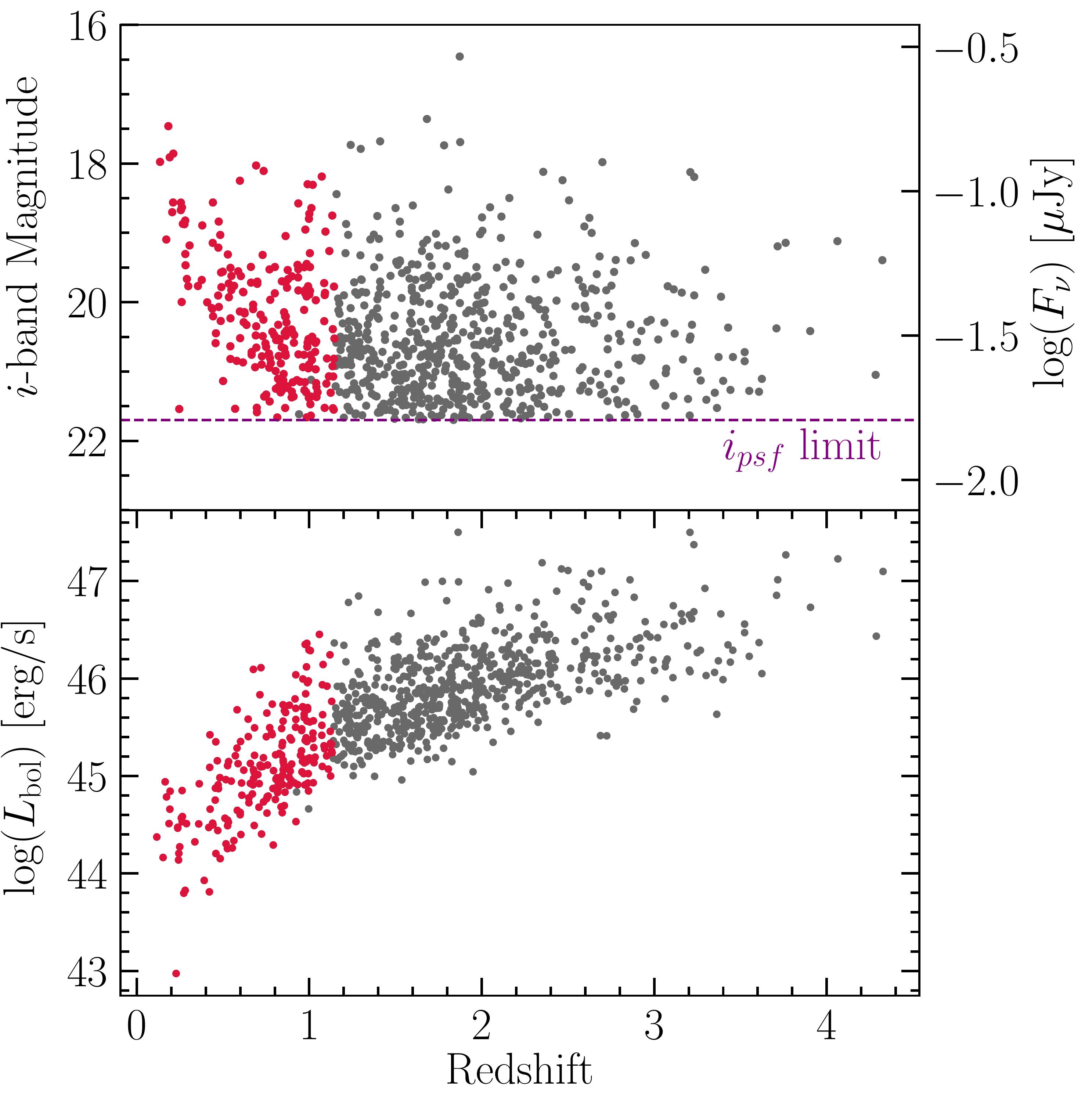}
\caption{\textbf{Top:} The $i$-band magnitude and redshift of the full SDSS-RM sample of 849 quasars (gray), along with the parent sample of 222, $z~<~1.13$ quasars used in this work (red). \textbf{Bottom:} The bolometric luminosity and redshift of the full SDSS-RM sample (gray) and $z~<~1.13$ sample used in this work (red). Bolometric luminosities are computed using monochromatic bolometric corrections of 9.26, 5.15, and 3.81 using the 5100$\rm \AA$, 3000$\rm \AA$ , and 1350$\rm \AA$ luminosities \citep{Richards2006}. Our SDSS-RM sample spans a broad range of luminosity and redshift and is more representative of the general quasar population than previous RM campaigns, see also Figure 1 of \citet{Shen2015}. \label{fig:full}}
\end{figure}

\begin{figure}
\includegraphics[scale = 0.18]{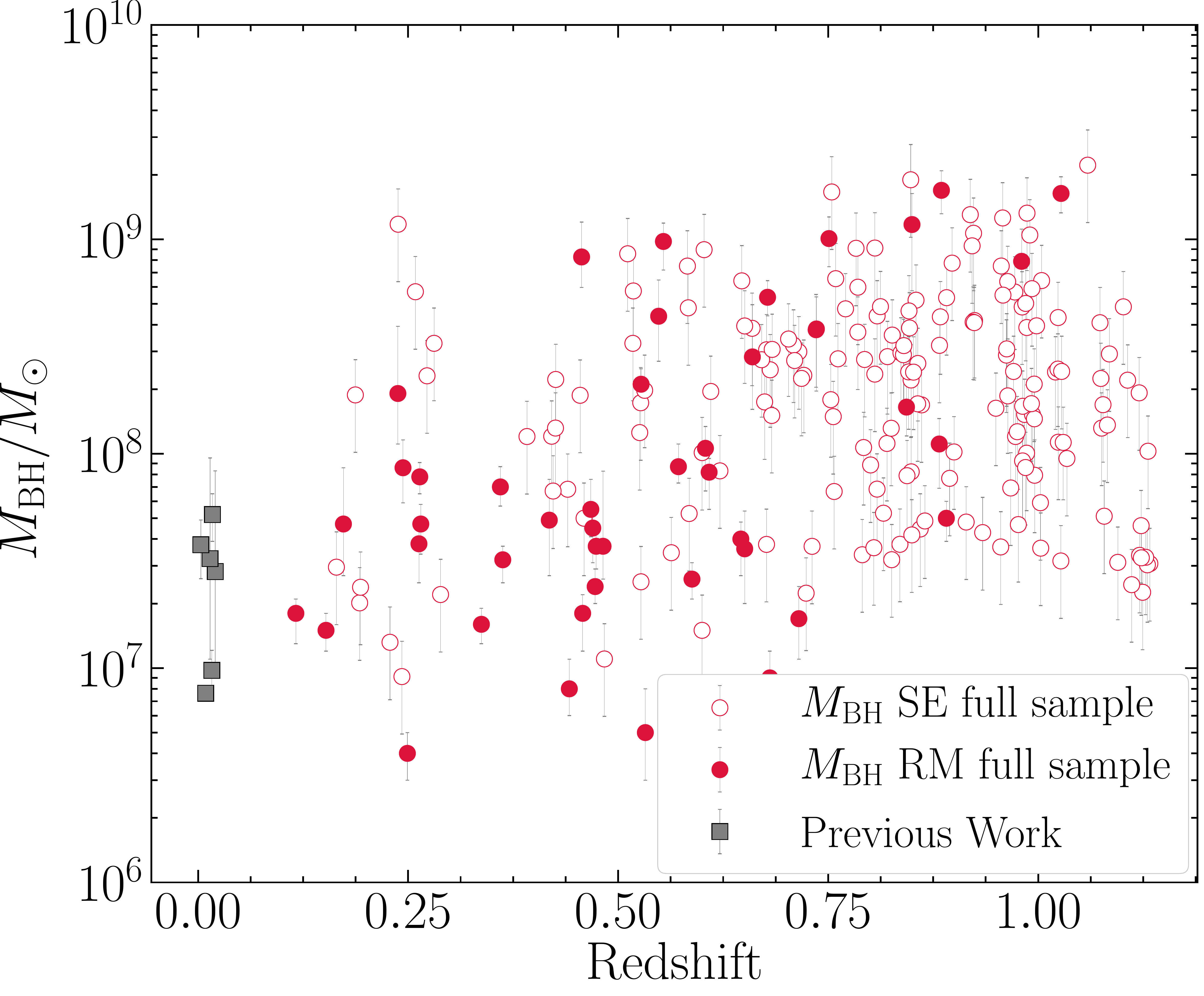}
\caption{The $M_{\rm BH}$ and redshift of our parent sample of 222 SDSS-RM quasars. Our sample is unique for accretion-disk RM as it has a large number of reliable black hole mass estimates: a total of 44 quasars in our sample have masses from broad-line RM \citep[red circles][]{Shen2016a,Grier2017}.  We supplement this data set with lower-precision single-epoch mass estimates for an additional 178 quasars (open symbols, from \citealp{Shen2016a} using the \citealp{Vestergaard2006} prescription). Filled squares show the limited number of previous measurements of both RM masses and accretion-disk sizes in broad-line AGNs for﻿NGC7469 \citep{Collier1998}, NGC 5548 \citep{Fausnaugh2016}, ﻿MCG +08-11-011 and NGC 2617 \citep{Fausnaugh2018} and NGC 4151 \citep{Edelson2017, McHardy2018} NGC 4395 and NGC 4593 \citep{McHardy2018} (NGC 4395 also has continuum RM measurements and a black hole mass from broad-line RM, but its $M_{\rm BH}$ of  $2 \times 10^5 M_{\odot}$ falls outside the figure). \label{fig:sub}}
\end{figure}

We here select the 222 quasars in SDSS-RM (see Figures \ref{fig:full} and \ref{fig:sub}) with $z~<1~.13$ previously studied for broad-line RM and black hole mass, $M_{\rm BH}$, estimates \citet{Grier2017}.  Of the 222 quasars, 44 have reliable $M_{\rm BH}$ estimates from \citet{Grier2017}, enabling us to study the accretion-disk structure dependence on black hole mass.
The selected sample is unique since it has well-measured BH masses and is suitable to study accretion-disk properties based on continuum lag measurements.

\subsection{Spectroscopy}
We use the Baryon Oscillation Spectroscopic Survey (BOSS) spectrograph \citep{Dawson2013, Smee2013} covering wavelengths of $3650-10400 $ \AA with a spectral resolution of $R \sim 2000$, with the spectrograph is mounted on the 2.5 m SDSS telescope \citep{Gunn2006}. Our study, uses the first year of SDSS-RM spectroscopic observations, obtained during seven dark/grey observing windows in Jan - Jul 2014. Each epoch has a typical depth of $S/N^2_g > 20$ (the average extinction-corrected $S/N^2$ per pixel in $g$ band evaluated at $g_{\rm psf} = 21.2$) \citep{Shen2015}, with a total of 32 spectroscopic epochs separated by a median of 4 days, with varying cadence depending on weather conditions and scheduling constraints. 

The spectroscopic data processing is initially processed using the standard SDSS pipeline \citep{Bolton2012} for flat-fielding, 1d extraction, wavelength calibration and a first pass at sky subtraction and flux calibration. SDSS-RM data are also processed with a second round of sky subtraction and flux calibration using a custom pipeline that uses position-dependent calibration vectors \citep[see][]{Shen2015} for details. Finally, a software package called \texttt{PrepSpec} is used to model the spectra and remove any remaining epoch-dependent calibration errors. This step is implemented by fitting a simple model for quasar spectra and considering a wavelength-dependent and time-dependent component to the continuum and a non-variable component to the narrow emission line fluxes. See \citet{Shen2016a} for details.  
 
We measure synthetic photometry in the $g$ and $i$-bands by integrating the SED with the SDSS filter response function \citep{Fukugita1996, Doi2010} and the flux errors. The synthetic flux error is computed using the quadratic sum of errors in the measured SED, errors in the shape of the response function and the errors in \texttt{PrepSpec} calibration.

Following \citet{Grier2017} we excluded epoch 7 (MJD = 56713) out of the 32 available epochs because it was taken under poor observing conditions, had significantly lower S/N, and was frequently ($>>$1/3 of the time) a $>$1$\sigma$ outlier compared to the other epochs. Furthermore, to improve the overall quality of the obtained continuum light curves, a small number of epochs ($1\%$) are rejected as outliers if offset from the median flux by more than five times the normalized median absolute deviation (NMAD), this is implemented to mostly remove data points where the fibers were incorrectly placed altering the flux or dropped fibers.
\subsection{Photometry}\label{sec:photometry}
SDSS-RM is supported by ground-based photometry from the 3.6m Canada-France-Hawaii Telescope (CFHT) and the 2.5m Steward Observatory Bok telescope. Between Jan and Jun 2014 the Bok/90 Prime instrument \citep{Williams2004} obtained 31 epochs in \textit{g}-band and 27 epochs in \textit{i}-band during 60 observing nights in bright time. The CFHT MegaCam \citep{Aune2003} obtained 26 epochs in \textit{g} and 20 epochs in \textit{i}-band. 

The photometric light curves are computed using image subtraction as implemented in the ISIS package \citep{Alard2000}. ISIS first creates a reference image using the best seeing exposure, then matches the astrometry of subsequent frames with different point-spread functions (PSF). This step uses a least-squares fit to find the optimal kernel between the reference image and the target image while accounting for PSF variation in each target image. The target image is then convolved and subtracted from the reference image to produce the light curves. The reference image and image subtraction is performed for each individual telescope, filter, CCD and field (Kinemuchi et al. 2018).

\subsection{Light Curve Merging}\label{lightmerging}

The combined monitoring from the SDSS, Bok, and CFHT telescopes provide a total of 88 epochs of $g$-band photometry and 78 epochs of $i$-band photometry. The mean fractional variability is 8.4\% in the $g$-band and 7.3\% in the $i$-band, in both cases calculated as the maximum-likelihood intrinsic variability accounting for observational uncertainties (following \citealp{Almaini2000,Sun2015}).
However, combining the three light curves is nontrivial, since each observatory has different seeing conditions and calibration issues for each filter response, telescope throughput and  and any other site-dependent calibration. We use the \texttt{CREAM} software \citep[Continuum REprocessing AGN Markov Chain Monte Carlo;][]{Starkey2016} to inter-calibrate the lightcurves obtained at different sites with the following model:
\begin{equation}
F_\nu(t) = \bar{F} + \Delta F\, X(t)
\ ,
\end{equation}
where the lightcurve shape $X(t)$ is normalized to $\left<X\right>=0$ and $\left<X^2\right>=1$ so that $\bar{F}(\lambda)$ is the mean and $\Delta F(\lambda)$ is the rms flux of the lightcurve. CREAM uses a power-law prior on the power spectrum of $X(t)$ (see Equation 8, 9 and 10 of \citealp{Starkey2017}) so that $X(t)$ by default resembles the observed behavior of AGN lightcurves (see \citealp{Grier2017} for step-by-step description of the lightcurve merging procedure). The fit allows $\bar{F}$ and $\Delta F$ to be different for the data from each site, while applying the same $X(t)$ to all sites. The site-to-site differences in $\bar{F}$ and $\Delta F$ then allow the data from each site to be scaled and shifted
and thereby effectively merged into a single lightcurve dataset with a common photometric calibration. This was done independently for the $i$ and $g$ photometry, thus defining a (slightly) different $X(t)$ for each band.

\begin{figure}
\includegraphics[scale = 0.35]{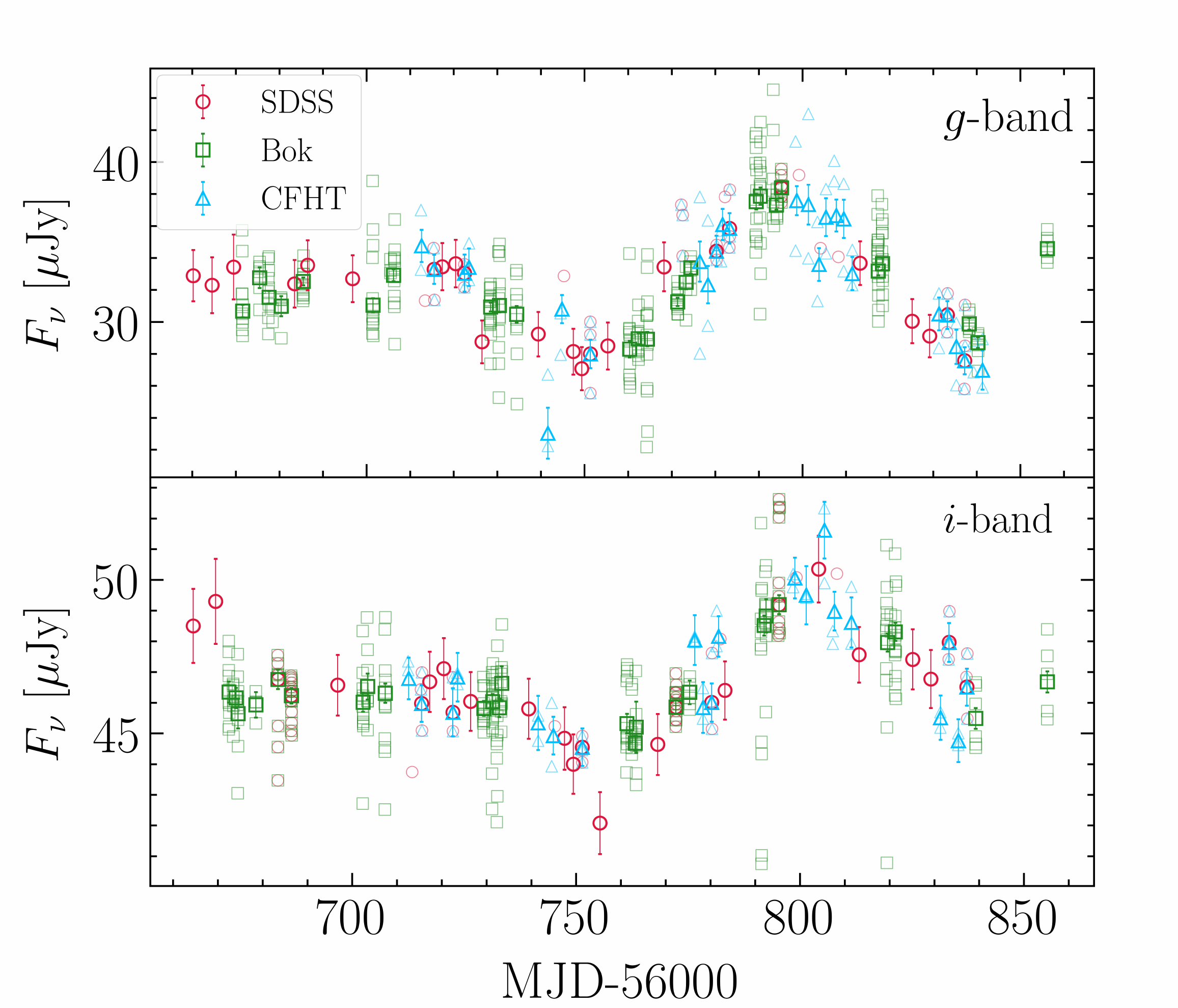}
\caption{Merged $g$ and $i$-band lightcurves for RM 267 as an example of the cadence and quality of our photometry.  Different symbols and colors indicate data from Bok (green) and CFHT (blue) photometry and SDSS (red) spectroscopy.  Bold symbols indicate nightly averages of the individual observations shown by fainter symbols. Our quasars have a total of 88 epochs in $g$ and 78 epochs in $i$ spanning a total of about 180 observed-frame days. \label{lciccf}}
\end{figure} 
\section{Continuum RM Analysis}
The SDSS-RM light curves are irregularly-sampled due to weather conditions and constraints on telescope allotted time; thus the RM analysis requires interpolation between epochs. We use two approaches to interpolate and measure lags and uncertainties from the merged light curves.
\subsection{ICCF}\label{iccf}
Our first RM analysis methodology is the Interpolated Cross Correlation Function \citep[\texttt{ICCF};][]{Gaskell1986, Gaskell1987, White1994, Peterson2004} where observations from different epochs are linearly interpolated to create an evenly sampled grid and calculate the Pearson coefficient $r$ between the two mean-subtracted light curves $S_1(t)$ and $S_2(t)$. The first light curve is then shifted by a time lag $\tau$ and $r$ is re-measured. This step is repeated across the range of allowed $\tau$, thus constructing the cross correlation function. The same procedure is repeated by shifting the other light curve by all $\tau$ values, and the final correlation function is averaged between the two.
Determining well-measured lags using the \texttt{ICCF} method is challenging considering the correlated errors associated with the lightcurve interpolation. We estimate errors on the \texttt{ICCF} lags using Monte Carlo (MC) iterations for flux resampling and random subset selection \citep{Peterson2004}, implemented using the publicly available \texttt{PyCCF} software \citep*{Sun2018b}. The flux in each point is resampled by a Gaussian distribution determined by its uncertainty, a random subset of epochs is chosen (with replacement), and the lag is recomputed. Repeated MC is used to obtain cross-correlation peak distribution (CCPD). The centroid of the CCF is restricted to the region where the CCF is above $80\%$ fraction of the peak; experimentation reveals that using the centroid of the CCF  rather than the CCF peak results in less biased lags and yields higher precision in virial masses \citep{Peterson2004}, we thus choose to work with cross-correlation centroid distribution (CCCD).

We adopt a delay grid spanning $\pm100$ days with spacing of half the mean of minimum separation between observed epochs. This search baseline is roughly half the total 180-day range of the SDSS-RM observations, and effectively prevents matching non-overlapping features between the light curves. We perform 5000 MC iterations over the range of allowed $\tau$ per light curve, returning the CCCD for the lag centroid $\tau_{cent}$ and the cross-correlation Pearson coefficient $r$ at each time delay within the the range. 

Each of the \texttt{ICCF} MC realizations is tested for correlation coefficient and significance of the lag and returns a ``failed peak" if significance criteria are not met (i.e., CCF peak is found to be on the upper or lower limit of the delay grid or if the correlation coefficient is less than 0.2 for data points within the centroid). Out of the unique sample of 222 RM objects, RM173 showed the most failed peak detection with only 37 successful detected peaks out of 5000 MC realizations. We therefore exclude this quasar as its CCCD is not statistically significant (We will shortly see that JAVELIN is also unable to obtain the continuum model for RM 173). In the rest of our sample $\sim$ 30$\%$ of objects have all 5000 successful MC realizations and on average each object has $\sim$ 85$\%$ success rate.
\subsection{JAVELIN}\label{javelin}
We also compute lags using the \texttt{JAVELIN} software \citep{Zu2011}. JAVELIN assumes a damped random walk (DRW) model to predict the lightcurves at unmeasured times. Observations confirm that the DRW model is a reasonable first-order description of quasar light curve variability on timescales of $\gtrsim$1~day, with variability amplitude and damping timescale\footnote{The typical damping timescale of a quasar in observed-frame is $\sim$ 1500 days \citep{Kelly2009, MacLeod2012}. Since our monitoring duration is shorter than the DRW damping timescale, our light curves are essentially modeled as a red-noise random walk with no damping. We explicitly tested damping timescales of 200-2000~days and found no significant differences in the best-fit \texttt{JAVELIN} lags.} dependent on quasar luminosity \citep{Kelly2009,Koz2010,MacLeod2010,Sun2018d}. The DRW in the continuum is first modeled by two priors to compute the continuum light curve variability with the assumption of covariance between times $t_i$ and $t_j$: 
\begin{equation}
<S_1(t_i)S_1(t_j)> \, = \,\sigma^2\, \left( 1-e^{-|t_i-t_j|/\tau_{\rm d}} \right)
\end{equation}
Here $\tau_d$ is the damping timescale.
This variability model can be approximated as a double-power law, with a short  timescale ($\Delta t < \tau_d$ rms of $\sigma \sqrt{2\,\Delta t/\tau_d}$ (power spectrum power-law of $\alpha=-2$) and a long timescale rms of $\sigma$ ($\alpha=0$).

\texttt{JAVELIN} models the reverberation response $\Psi(\tau)$ as a top-hat function centered at $\bar{\tau}$ with full width $\Delta\tau$. The reverberating light curve is then the ``lagged" version of the driving light curve smoothed and scaled by the parameters of the top-hat function. 

\texttt{JAVELIN} uses a two-step Markov Chain (MCMC) simulation \citep{Zu2011}. The first step analyzes the driving light curve by itself and obtains uncertainties and posterior distributions for the DRW parameters $\tau_d$ and $\sigma$. The second MCMC analysis determines the best-fit transfer function centroid $\bar{\tau}$ and $\Delta \tau$ based on the posterior distribution from the isolated continuum in the first MCMC, where each DRW parameter is the median value with the Gaussian width chosen to match the upper and lower 1$\sigma$ confidence regions.
This approach results in three new posteriors: mean lag $\bar{\tau} = (\tau_i+\tau_j)/2$, the width of the top-hat $\rm \Delta \tau = \tau_j - \tau_i$, and a scaling coefficient $A$. The second MCMC process also updates the posterior distribution for the DRW parameters $\tau_d$ and $\sigma$. \texttt{JAVELIN} is able to allow for all the parameters of the DRW model and transfer function to vary in the MCMC; however, we fix the damping time scale $\tau_d = 200$~days. The assumed damping timescale does not affect the model light curves so long as it is longer than our 180-day monitoring duration. 
We similarly fix the transfer function to have a width of $\Delta\tau = 0.5$~day, after testing values between 0.25 and 1 day and finding no significant differences in the measured lags. The top-hat function used by \texttt{JAVELIN} is a simplification of the more complicated transfer function likely to describe accretion-disk reprocessing (see \citealt{Starkey2016}), but is a reasonable approximation so long as the disk response is short compared to the lag, and it has been the common assumption of previous work to which we make comparisons. The uncertainty of the DRW parameters is obtained based on the statistical confidence limits from the posterior distribution. 
\texttt{JAVELIN} fails to compute the continuum model for the RM 173 just as the \texttt{ICCF} failed, and also fails to compute the continuum model for RM 187 and RM 846. In the end we have 219 quasars that have computed \texttt{JAVELIN} lags. 

\begin{figure*}
\centering
\includegraphics[scale = 0.30]{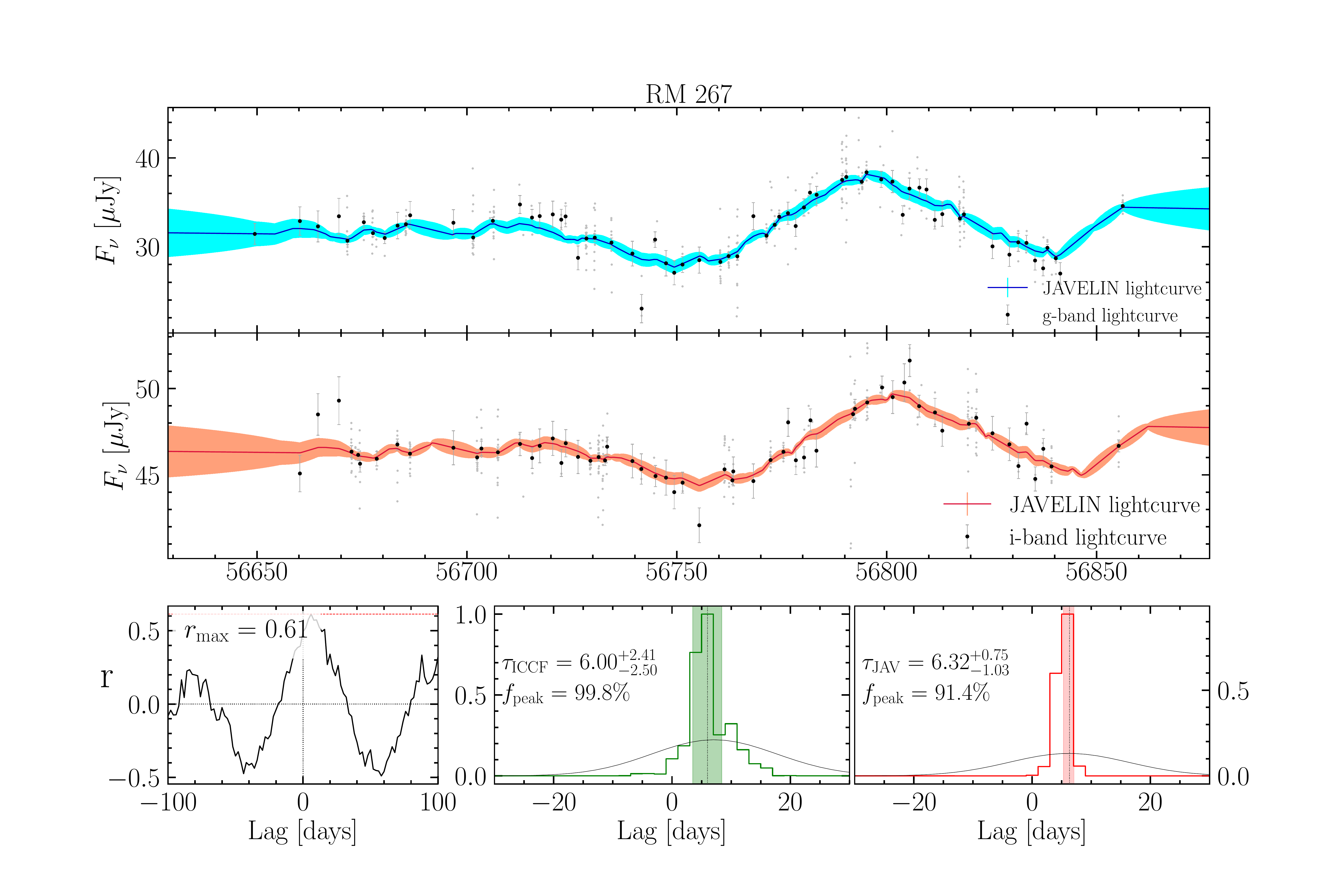}
\caption{\textbf{Top:}  Continuum $g$ (blue) and $i$-band (red) light curves and errors for quasar RM ID 267 computed with \texttt{JAVELIN}. For clarity, black points indicate the averages of data taken within a single night, although all lag analyses were performed on the individual, non-averaged observations displayed by small grey points. The best-fit \texttt{JAVELIN} DRW models are shown by the shaded lines in each panel. \textbf{Bottom left}: The cross correlation coefficient computed at each lag with its maximum identified by a red horizontal line. \textbf{Bottom center}: Lag probability distribution computed by \texttt{ICCF}, with the local minima of the primary peak indicated by gray shading, and the identified lag and $\pm$1$\sigma$ error indicated by the green dotted line and shading. \textbf{Bottom right} Lag probability distribution computed by \texttt{JAVELIN}. The main lag and its $\pm$1$\sigma$ error are represented by the red dotted line and shading. In both plots the Gaussian-smoothed curve represents the smoothed peak with 5-day standard deviation. The complete figure set (219 images) is available on the online journal.\label{lc}}
\end{figure*}

We demonstrate continuum lag analysis results in Figure \ref{lc} for RM 267 for the $g$ and $i$ band continuum model using \texttt{JAVELIN} and \texttt{ICCF}. Similar figures for our full sample are provided as a Figure set.  
\subsection{Lag Identification Method}\label{lag_ID}
Identifying a well-measured lag from the methods described in \ref{iccf} and \ref{javelin} requires additional checks to eliminate cases that appear to be unreliable or ambiguous. Additionally, in many cases the CCCDs obtained from our methods have multiple peaks that correspond to aliases in the lags due to semi-repeating features in the light curves. Also, it is not always clear if the initial reported lag corresponds to genuine reverberation. We devise a set of criteria to identify unambiguous lags, likely to correspond to real reverberation, while rejecting less reliable lags.

\subsubsection{Alias Removal}\label{sec:alias}
As mentioned above, many of our quasars have CCCDs with multiple peaks, corresponding to competing alternatives for the CCF lag. Some of these peaks occur at the bounds of the time window ($\pm$ 100 days) and are caused by numerical issues. 

We assume a prior that lags are most likely to be detected when the two light curves have maximal overlap. Conversely, if shifting epochs by a time delay results in zero overlapping data points between common epochs then the probability of finding a lag will be zero. We adopt the same weighting and alias removal scheme as in \citet{Grier2017}. The weight is defined as $P(\tau) = (N(\tau)/N_{0})^2$ ; with $N(\tau)$ corresponding to the number of overlapping epochs between the $g$ light curve and the $i$ light curve shifted by lag $\tau$, and $N_0$ corresponding to the maximum number of overlapping epochs from \textit{g} and \textit{i} light curves at zero time delay $\tau = 0$.

\begin{figure}
\centering
\includegraphics[scale = 0.38]{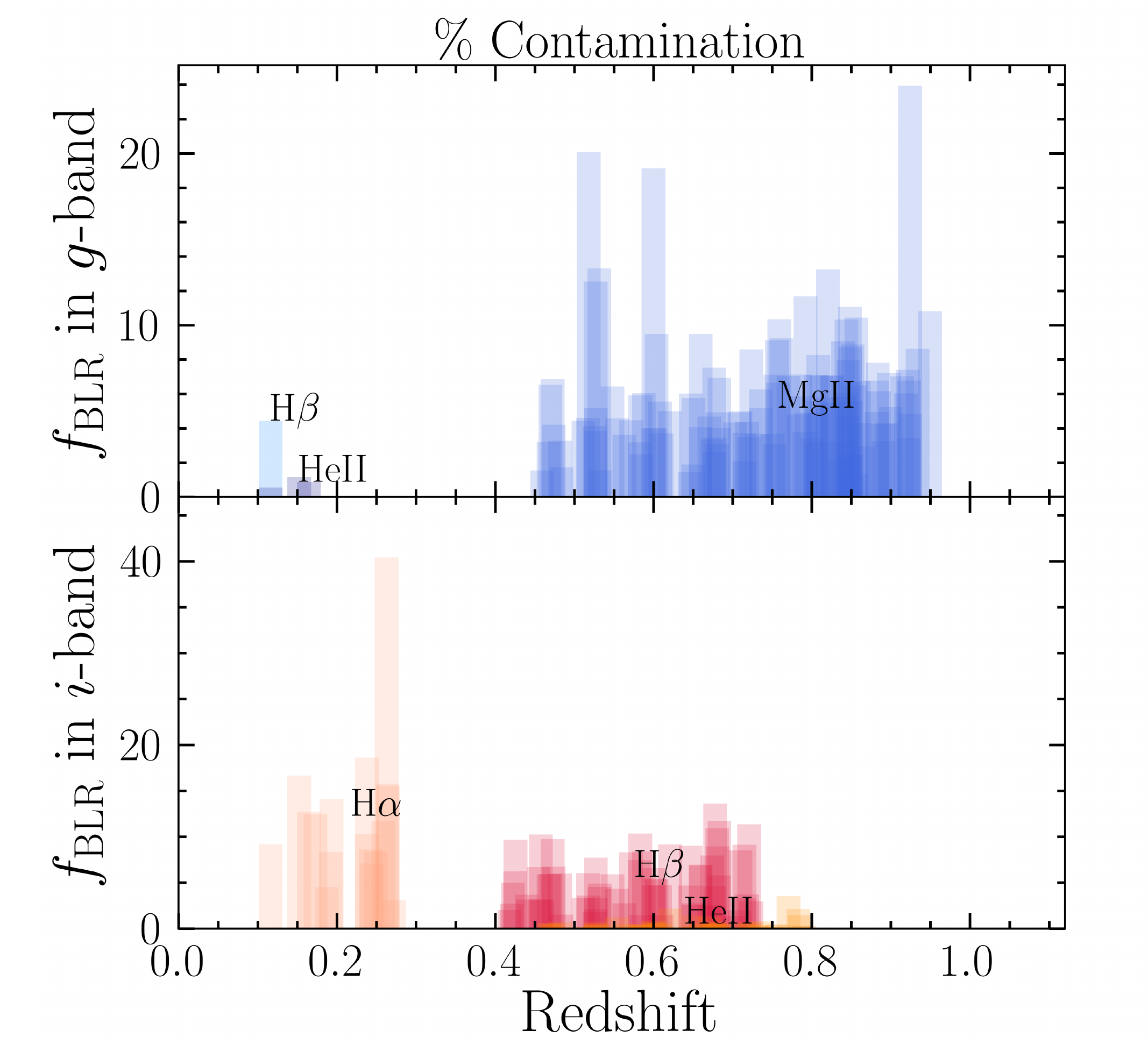}
\caption{Contamination by different broad emission lines in the $g$-band (top) and $i$-band (bottom) photometry of our 222 quasars obtained from Shen et al. 2018 (in prep). Broad-line contamination, $f_{\rm BLR}$, is calculated as EW(line)~/~FWHM(band). We require $<$12.5\% broad-line contamination for a ``well-defined'' photometric accretion-disk lag. As shown in the bottom panel of Figure~\ref{fig:javdiagnostic} few of the quasars have more than 12.5\% maximum contamination in $g$ and $i$ band.\label{fig:BLR_contamination}}
\end{figure}

Our general framework for finding lags is based on \texttt{JAVELIN} posterior distribution as CCCD.
The CCCD is weighted by $P(\tau)$ to avoid alias lag solutions and smoothed using a Gaussian filter with a width of five days. The smoothing is used to identify peaks in the weighted CCCD as well as the local minima around each peak. The weighted, smoothed CCCD may contain multiple peaks with a high-significance peak accompanied by multiple low-significance peaks. We compute the area between consecutive local minima and identify the local minima that contain the peak with the most area and adopt the lag as the median of the un-smoothed CCCD within the identified local minima.
Furthermore, this technique is helpful in identifying more plausible lags for those CCCDs that show peaks on either ends of the lag interval. 

The lag uncertainty is computed as the mean absolute deviation relative to the median, computed between the local minima on either side of the peak.

\subsubsection{BLR impact on Continuum Light Curves}\label{sec:BLR}
The $g$ and $i$ photometric bands in our lightcurves may include substantial flux from broad emission lines in addition to the continuum emission. Considering that BLR lags typically have longer timescales and show smaller-amplitude variability compared to continuum lags \citep{MacLeod2012}, BLR contamination may potentially affect the observed time lag derived from the continuum. We consider emission lines that could fall in range of SDSS filters depending on the redshift of our quasar sample: CIV , CIII, MgII, H$\beta$ and H$\alpha$ at respectively 1550, 1909, 2799, 4861, 6563 \AA ~in the rest frame. We determined the broad-line contribution, $f_{\rm BLR}$ in each as the ratio of emission-line equivalent width \citep{Shen2018} to the SDSS filter effective width \citep{Fukugita1996}. The contamination result for all of the objects in our sample is illustrated in Figure \ref{fig:BLR_contamination}.

\subsection{Criteria}\label{criterion}
We require additional tests to identify if our computed lag are statistically significant. One of the tools on which we rely is the maximum cross correlation coefficient, $r_{\rm max}$, as a measure of correlation between the \textit{g} and \textit{i} light curves. Visual inspection of the \textit{g} and \textit{i} light curves and computed lag probability distributions revealed that a threshold of $r_{\rm max} > 0.4$ can eliminate non-correlated light curves. Another tool  used to identify the significance of the main peak is the fraction of the probability distribution that lies within the primary peak, hereafter referred to as ``$\rm f_{\rm peak}$". We define $\rm f_{\rm peak}$ as the ratio of the weighted CCCD between the local minima,  used in the lag calculation to the the prior-weighted CCCD across the full $\pm 100$ day delay range. We accept only peaks that carry more than $75\%$ of the total posterior probability ($f_{\rm peak} >$ 0.75) to obtain a sample of well-measured lags from our quasar sample. We also want to avoid lags that are contaminated by BLR emission lines, as discussed above in section \ref{sec:BLR}. We thus exclude any objects with emission-line contaminations greater that 12.5\%.

In summary, our criteria for accepting a lag as ``well-defined" lags are as follows:
\begin{itemize}
\setlength{\itemsep}{1pt}
\item{$r_{\rm max} >$ 0.4 : Minimum cross-correlation to consider that corresponds to physical reverberation}
\item{$f_{\rm peak} >$ 75\%: Threshold to reject ambiguous lags with significant support for competing aliases}
\item{$f_{\rm BLR} <$ 12.5\%: Minimal broad-line contribution in both $g$ and $i$ photometric light curves}
\end{itemize}
\begin{figure}[!h]
\includegraphics[scale = 0.28]{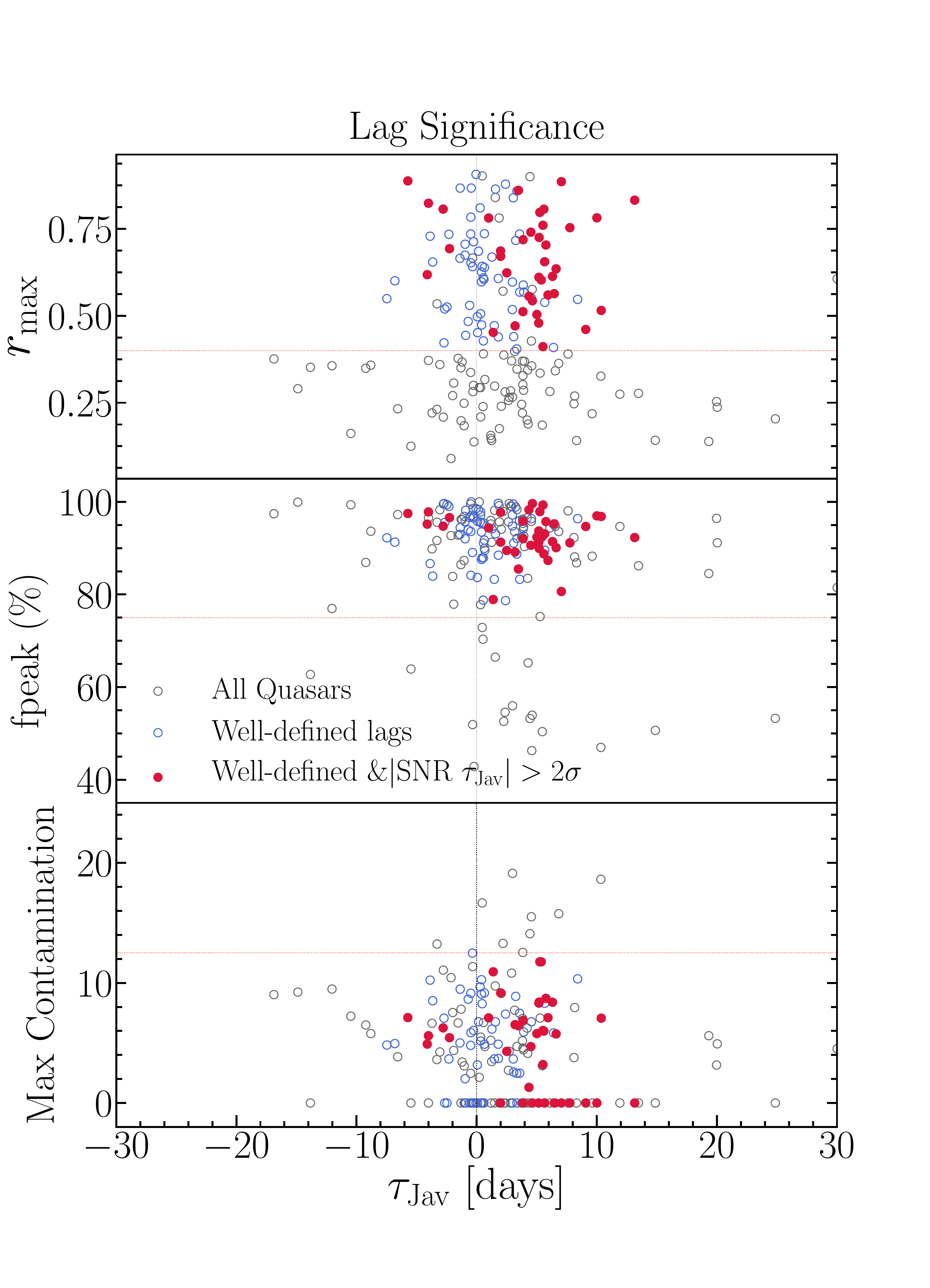}
\caption{Our three criteria for ``well-defined" lags versus the computed \texttt{JAVELIN} lags for the our sample of 222 quasars (gray symbols).  Quasars with ``well-defined" lags meeting our criteria are shown in open blue symbols and the ``high-SNR" lags that are 2$\sigma$ significant are shown in red.  \textbf{Top}: Maximum cross-correlation coefficient $r_{\rm max}$ from the $g$ and $i$-band light curves. The horizontal red dotted line indicates the minimum $r_{\rm max}>0.4$ criterion required for a ``well-measured" lag. \textbf{Middle:} Fraction $f_{\rm peak}$ of the probability distribution that lies within the primary peak, where the horizontal red dotted line represents the minimum $f_{\rm peak}>75\%$ ``well-defined" lag criterion.  \textbf{Bottom:} Maximum broad-line contamination in each of  $g$ and $i$ bands.  The dotted red horizontal line indicates the maximum allowed broad-line contamination for a ``well-defined" lag, $f_{\rm BLR}<12.5\%$. \label{fig:javdiagnostic}}
\end{figure}

Our final lag sample is reported in Table \ref{table1} for the first 10 of all the 95 quasars that satisfy the above criteria. We also report redshifts \citep{Shen2015}, RM $M_{\rm BH}$ and single-epoch $M_{\rm BH}$ from \citet{Grier2017},  $\lambda L_{\lambda 3000}$ \citep{Shen2015}, and the observed-frame lag and uncertainties using both \texttt{ICCF} and \texttt{JAVELIN}.

\section{Lag Reliability}\label{sec:reliability}
The \texttt{JAVELIN} method produces a total of 95 ``well-defined" lags that satisfy the reliability criteria defined in section \ref{criterion}. From the ``well-defined" sample of 95 continuum lags, we also construct a subsample of 33 ``high-SNR" lags that are 2$\sigma$ different from zero;  ${\rm SNR}(\tau_{\rm JAV})~\geqslant ~ 2$ in addition to meeting the criteria listed in Section \ref{criterion}. Summarizing, we use the following definitions for our main sample of ``well-defined" lags and the subsample of ``high-SNR" lags:
\begin{itemize}
\item{``well-defined" lags: $r_{\rm max} >$ 0.4, $f_{\rm peak} >$ 75\% and $f_{\rm BLR} <$ 12.5\%}
\item{``high-SNR" lags:  $r_{\rm max} >$ 0.4, $f_{\rm peak} >$ 75\%, $f_{\rm BLR} <$ 12.5\% and ${\rm SNR}(\tau_{\rm JAV})\geq 2$} 
\end{itemize}

Due to the limits in the SDSS-RM survey our measured lags could impose selection bias: For example the ``high-SNR" lag sample includes only larger lags while the ``well-defined" lag sample may be more representative of the broader quasars population. We will discuss this point in more detail in Appendix A.

One of the difficulties in reverberation mapping, particularly for monitoring surveys such as SDSS-RM, with relatively sparse cadence and non-negligible flux uncertainties, is knowing if there is genuine reverberation rather than a false detection caused by a chance similarity between light curves. Chance similarities would create equal number of positive and negative lags, while reverberation would produce only positive lags, with some negative lags due to noise or sampling properties of light curves. We investigate this issue with set of plots presented in Figure \ref{fig:javdiagnostic}. Our lag-finding analysis and ``well-defined" lag criteria include no explicit or implicit preference for a positive lag from \textit{g} to \textit{i}-band.  The high-SNR sample has 33 positive lags and only 5 negative \texttt{JAVELIN} lags, indicating that most objects have genuine reverberation with a false positive rate (i.e., ratio of negative to positive lag) of only 15\%. The ``well-defined" lag sample has 68 positive lags with 27 negative lags, similarly showing a significant excess of positive lags. The larger number of negative lags in the ``well-defined" sample is expected from the broad lag CCCDs of many of the quasars. 

\begin{figure}
\includegraphics[scale = 0.24]{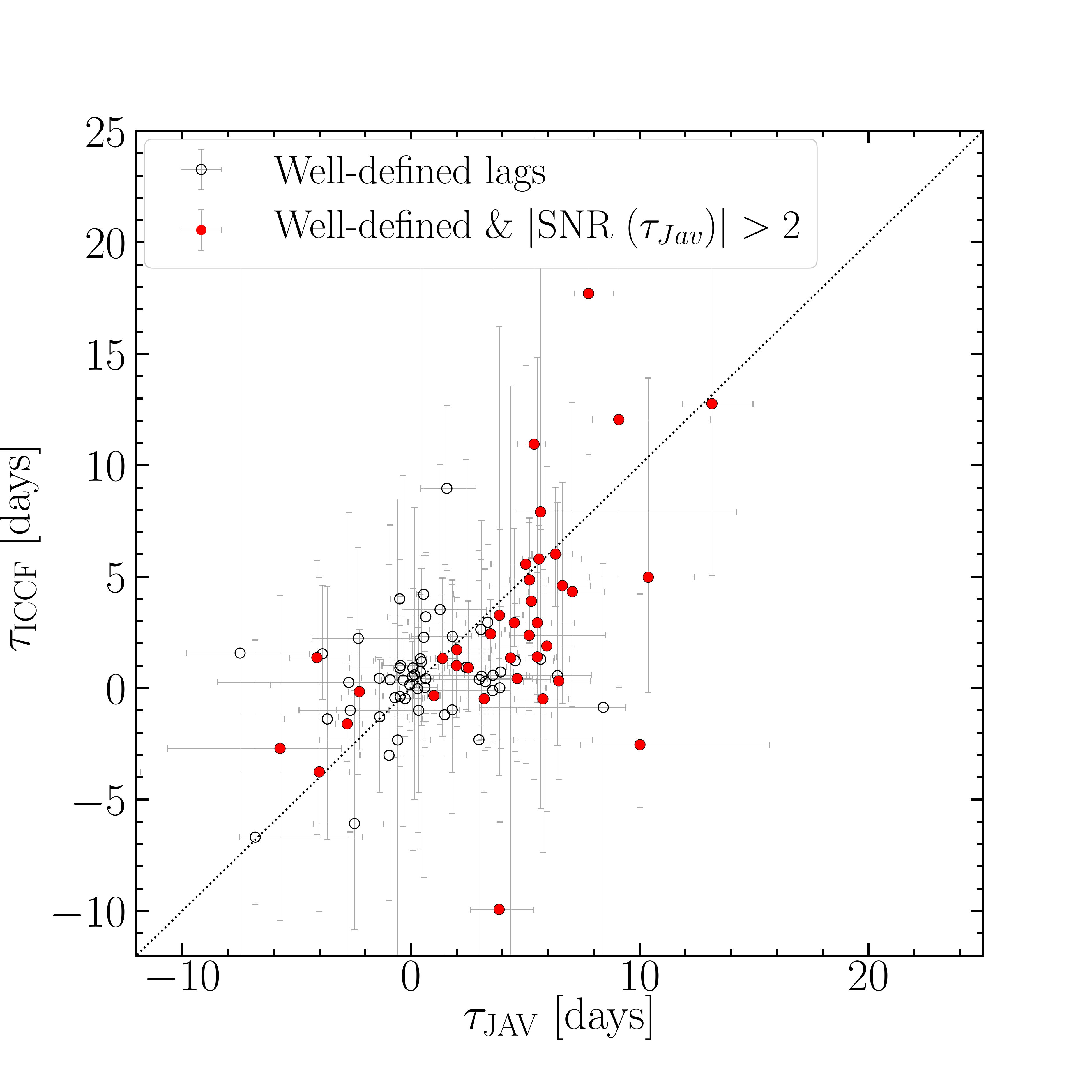}
\caption{\texttt{ICCF} vs \texttt{JAVELIN} lags from our sample of ``well-defined" \texttt{JAVELIN} lags that meet the criteria outlined in Section \ref{criterion} (rmax$>$0.4, fpeak$>$0.75, ${\rm f_{BL}}<$0.125). Lags that additionally have $|$SNR$(\tau)|>$2 are illustrated by red solid circles.  The \texttt{ICCF} and \texttt{JAVELIN} methods find consistent lags, although the \texttt{ICCF}	 method generally has larger error bars due to its (simplistic and unrealistic) assumption of linear interpolation between measured fluxes.\label{fig:strong_weak}}
\end{figure}

We compare our two lag methodologies, \texttt{ICCF} and \texttt{JAVELIN}, in Figure \ref{fig:strong_weak}. Most sources have differences between their \texttt{ICCF} and \texttt{JAVELIN} lags indicating that the \texttt{ICCF} uncertainties are over estimated; $<|(\tau_{jav} - \tau_{iccf})/\sigma_{jav}|>$ = 1.29 and  $<|(\tau_{jav} - \tau_{iccf})/\sigma_{iccf}|>$ = 0.41. When comparing the two methodologies, we note that \texttt{JAVELIN} presents an empirically motivated model for interpolating the light curve by explicitly assuming that the power spectral density is a DRW model, while implicitly assuming a prior that the two light curves are reverberating. \texttt{ICCF} does not make this assumption, and instead linearly interpolates between measurements to describe the light curve. The broad agreement between \texttt{JAVELIN} and \texttt{ICCF} lags is expected given our relatively short ~4-day cadence and low quasar variability observed on short timescales (e.g. \citealt{MacLeod2012,Mushotzky2011}). Simulations also indicate that \texttt{JAVELIN} and \texttt{ICCF} find similar and consistently reliable lags (\citealt{Zu2011}; Li et al. 2019 in prep.) Appendix B additionally tests the effects of unmeasured variability between the observational cadence, and finds that both \texttt{JAVELIN} and \texttt{ICCF} return statistically consistent lags even if we assume implausible large short timescale variability.

Visually inspecting the \texttt{ICCF} and \texttt{JAVELIN} results shows that the two methods generally identify consistent lags, although the computed uncertainties in the \texttt{ICCF} method are larger than \texttt{JAVELIN}. Figure \ref{fig:strong_weak} illustrates the general consistency in lag measurements between the two methods, suggesting that \texttt{JAVELIN}'s model is not introducing any unknown biases into our measurements that are not also inherent to the \texttt{ICCF} method. 

\startlongtable
\begin{deluxetable*}{ccccccccc}
\tablecaption{``Well-defined" quasar sample information\label{table1}}
\tablehead{
\colhead{RMID} & \colhead{RA} & \colhead{Dec} & \colhead{z} & \colhead{log $M_{\rm BH}$} & \colhead{log $\lambda L_{\rm \lambda 3000}$} &  \colhead{$\tau_{\rm ICCF}$} & \colhead{$\tau_{\rm JAV}$} & \colhead{SNR($\tau_{\rm JAV}$)} \\
\colhead{} & \colhead{(deg)} & \colhead{(deg)} & \colhead{} & \colhead{($M_{\odot}$)\tablenotemark{a}} & \colhead{(erg $s^{-1}$)} & \colhead{(days)} & \colhead{(days)} & \colhead{\tablenotemark{b}}
}
\startdata
\hline
016 & 214.0290 & 53.1583 & 0.848 & 	$9.07_{-0.26}^{+0.22}$ 	& 	44.85 	& 	$-3.76_{-6.26}^{+8.74}$   & 	$-4.01_{-7.82}^{+1.31}$	 & -3.07 \\ 
017 & 213.3511 & 53.0908 & 0.456 &  $8.92_{-0.19}^{+0.24}$  &   44.16   &   $2.93_{-3.21}^{+2.24}$    &     $5.52_{-1.68}^{+1.62}$   & 3.30 \\ 
029 & 213.2946 & 52.9640 & 0.816 &  $7.72^{\star}$        & 44.12     &   $-1.00_{-3.70}^{+5.30}$   &     $0.33_{-1.29}^{+1.79}$   & 0.26 \\ 
061 & 214.0000 & 52.7378 & 0.983 &  $8.18^{\star}$        & 44.44     &   $-2.54_{-2.82}^{+6.76}$   &     $10.01_{-2.60}^{+5.67}$  & 3.86 \\ 
062 & 213.5737 & 53.4697 & 0.808 &  $8.64^{\star}$        & 44.25     &   $1.18_{-2.85}^{+4.18}$    &     $0.46_{-1.67}^{+1.58}$   & 0.27 \\ 
078 & 212.9757 & 53.1887 & 0.581 &  $8.88^{\star}$        & 44.57     &   $-0.11_{-1.98}^{+2.49}$   &     $3.57_{-3.79}^{+0.79}$   & 0.94 \\ 
088 & 212.9657 & 52.8956 & 0.516 &  $8.51^{\star}$        & 44.25     &   $-0.47_{-1.72}^{+2.94}$   &     $-0.25_{-0.34}^{+0.74}$  & -0.34 \\ 
101 & 213.0592 & 53.4296 & 0.458 &  $7.26_{-0.19}^{+0.17}$  & 44.64     &   $1.54_{-2.06}^{+3.08}$    &     $-3.87_{-0.56}^{+5.17}$  & -0.75 \\ 
102 & 213.4708 & 52.5790 & 0.860 &  $8.23^{\star}$        & 45.01     &   $0.91_{-1.94}^{+3.00}$    &     $2.51_{-1.03}^{+0.73}$   & 2.44 \\ 
118 & 213.5533 & 52.5358 & 0.714 &  $8.48^{\star}$        & 45.12     &   $0.90_{-2.64}^{+2.92}$    &     $-0.48_{-0.28}^{+0.49}$  & -0.99 \\ 
\enddata
\tablenotetext{a}{Single epoch masses are identified by $\star$ and are assumed to have an error of 0.4 dex.}
\tablenotetext{b}{The SNR is calculated accounting for the \texttt{JAVELIN} lag sign, if the lag is positive the SNR is positive, if the lag is negative the SNR is negative.}
\tablecomments{Table 1 is published in its entirety in the machine-readable format. A portion is shown here for guidance regarding its form and content.}	
\end{deluxetable*}

There is one additional object, RM 769, that has a $> 3 \sigma$ difference between lags from \texttt{ICCF} and \texttt{JAVELIN}. It is the only object with an \texttt{ICCF} lag that has a ``well-defined" peak that differs by $>3\sigma$. While inspecting the RM 769 light curve we found that the DRW models from \texttt{JAVELIN} are heavily influenced by a few flux measurements that have significantly lower observational uncertainties than the rest of the light curve. We experimented and found that if we increase all the uncertainties in the light curve by 3\% the \texttt{JAVELIN} results change dramatically and become consistent with the \texttt{ICCF} lag. Due to this object's small error, and  more than 3$\sigma$ difference from \texttt{JAVELIN} lag estimate we reject this object from our sample.

Table \ref{table2} presents a comparison of our SDSS-RM study with other multi-object continuum lag surveys. Our study's largest advantage is the availability of spectroscopic RM observations and resulting $M_{\rm BH}$ measurements, enabling a comparison of disk size with black hole mass. Further comparison of our measured accretion-disk properties with previous work is presented in section \ref{sec:5.1}.

\startlongtable
\begin{deluxetable*}{ccccccc}
\tablecaption{Comparison with other multi-object continuum lag surveys\label{table2}}
\tablehead{\colhead{Survey} & \colhead{Lags\tablenotemark{a}} & \colhead{Epochs\tablenotemark{b}} & \colhead{Cadence} & \colhead{Duration} & \colhead{Bands} & \colhead{RM $M_{\rm BH}$\tablenotemark{c}}}
\startdata
\hline
Pan-STARRS & 39 & 373 & 3~day & 3.3 years &  g,r,i,z  & No \\ 
OzDES & 15 & 30 & 7~days & 1~season~(180 days) & g,r,i,z & No \\
SDSS-RM & 95 & 83 & 4~days & 1~season~(180 days) & g,i & Yes \\
\enddata
\tablenotetext{a}{The number of reported lags for the main sample in \citet{Jiang2017a,Mudd2017}.}
\tablenotetext{b}{Median number of total epochs per band.}
\tablenotetext{c}{Time-domain spectroscopic coverage available for RM $M_{\rm BH}$ measurements.}
\end{deluxetable*}

\section{Discussion}\label{sec:discussion}
The photometric lags measured from SDSS-RM can be employed to measure accretion-disk sizes across a wide range of quasar properties. We use the SS73 model as expressed in Equation \eqref{eq:SS73_theory}, as a starting framework, comparing our measured lags to the expectations of the analytic thin-disk model. 

We follow a Bayesian approach and fit accretion-disk parameters using the full set of ``well-defined" lags. Although many of these lags have large error bars and are consistent with zero, their distribution still carries valuable information.  Appendix A also represents results from fitting only the high-SNR lags, demonstrating that restricting to positive lags results in biased accretion-disk fits.

We use the Bayesian framework implemented in the software package PyMC3 \citep{Salvatier2016} \footnote{Probabilistic programming in Python using PyMC3 https://doi.org/10.7717/peerj-cs.55} to fit accretion disk parameters. To sample the posterior we provide disk parameter priors as a normal distribution centered at the expectation from SS73 model. We sample our MCMC fit with 40,000 steps, discard the first 20,000 steps as burn-in phase, and explicitly check the Gelman-Rubin statistics \citep{Gelman1992} for convergence diagnostic. All the lags are reported in the observed-frame (i.e., $\tau_{obs}$) as we account for the effects of wavelength redshift and time dilation in our analysis.
\subsection{Disk Normalization}\label{sec:5.1}
We start with the SS73 model presented in Equation \eqref{eq:SS73_theory} and compute each object's individual accretion disk size $\tau_0$ 
following the equation for the SS73 model observed-frame lag $\tau$:
\begin{equation}\label{eq:SS73_Fit}
\tau_{\rm SS73} = \tau_0\,(1+z)^{-1/3}\Big[(\frac{\lambda_i}{9000 \AA})^{4/3}-(\frac{\lambda_g}{9000 \AA})^{4/3}\Big]
\end{equation}
We normalize wavelength by $\lambda_0 = \lambda/9000$ \AA $\,$ because it was found to minimize the correlation between the best-fit $\tau_0$ and wavelength scaling $\beta$ in Section \ref{sec:beta}.

For simplicity, we refer to each of the measured $c\tau$ and model-predicted $c\tau_{\mathrm SS73}$ as a ``disk size." More precisely, these quantities are the relative distances corresponding to the differences between the characteristic lags from each waveband.

 The analytic disk normalization $\tau_0$ is equal to:
\begin{multline}\label{eq:normalization}
\tau_0 = \frac{1} {c} \Big( \frac{45\,G} {16\,\pi^6\,h_p\,c^2}\Big)^{1/3} X^{4/3} \, {(9000 \AA)}^{4/3} \\ \Big( \frac{C_{Bol}} {\eta\,c^2} \Big)^{1/3}\, M_{\rm BH}^{1/3}\, {\lambda L_{\lambda 3000}}^{1/3}
\end{multline}

Here $M_{\rm BH}$ represents the BH mass from RM \citep{Grier2017} and single epoch measurements \citep{Shen2016b}. When both RM and single-epoch masses are available for a quasar, we use the RM mass. We compute the $L_{\rm bol}$ using a bolometric luminosity correction $C_{bol} = 5.15$ from \citet{Richards2006} for $\lambda L_{\lambda 3000}$ as $L_{\rm bol} = C_{bol}\, \lambda L_{\lambda 3000}$. The quantity $X$ accounts for the relatively broad width of blackbody radiation causing the response at a given wavelength to arise from a range of radii in the disk, including smaller radii where the blackbody radiation is proportional to $T$ on the Rayleigh-Jeans tail of the blackbody emission, and larger radii where the increasing disk surface area is offset by the exponential Wien cutoff. Given a $T(r)$ profile and a wavelength $\lambda$, the observed mean 
delay is $\tau=r(\lambda)/c$ where $\lambda = hc / XkT(r(\lambda))$.  We follow previous work (\citealt{Fausnaugh2016,Mudd2017}) and calculate $X$ by assuming that $r(\lambda)$ is the flux-weighted mean radius for emission at $\lambda$ from a face-on disk of pure blackbody emission with $T(r) \propto r^{-3/4}$, which yields $X=2.49$.  
Comptonization and other radiative transfer effects may also affect the disk emission profile (e.g., \citealt{Davis2005,Slone2012}), potentially making $X$ a function of radius (or wavelength).  We adopt the global blackbody assumption of $X=2.49$ as a point of comparison for comparing to the SS73 model, noting that larger or smaller continuum lags may result from non-blackbody radiative transfer effects in addition to structural changes in the SS73 model.

In the following analysis we adopt $\eta$~=~0.1, $C_{bol}$~=~5.15 and X~=~2.49 when we plot the SS73 model in Figures 9, 12, 13, 14.

\begin{figure}
\includegraphics[scale = 0.22]{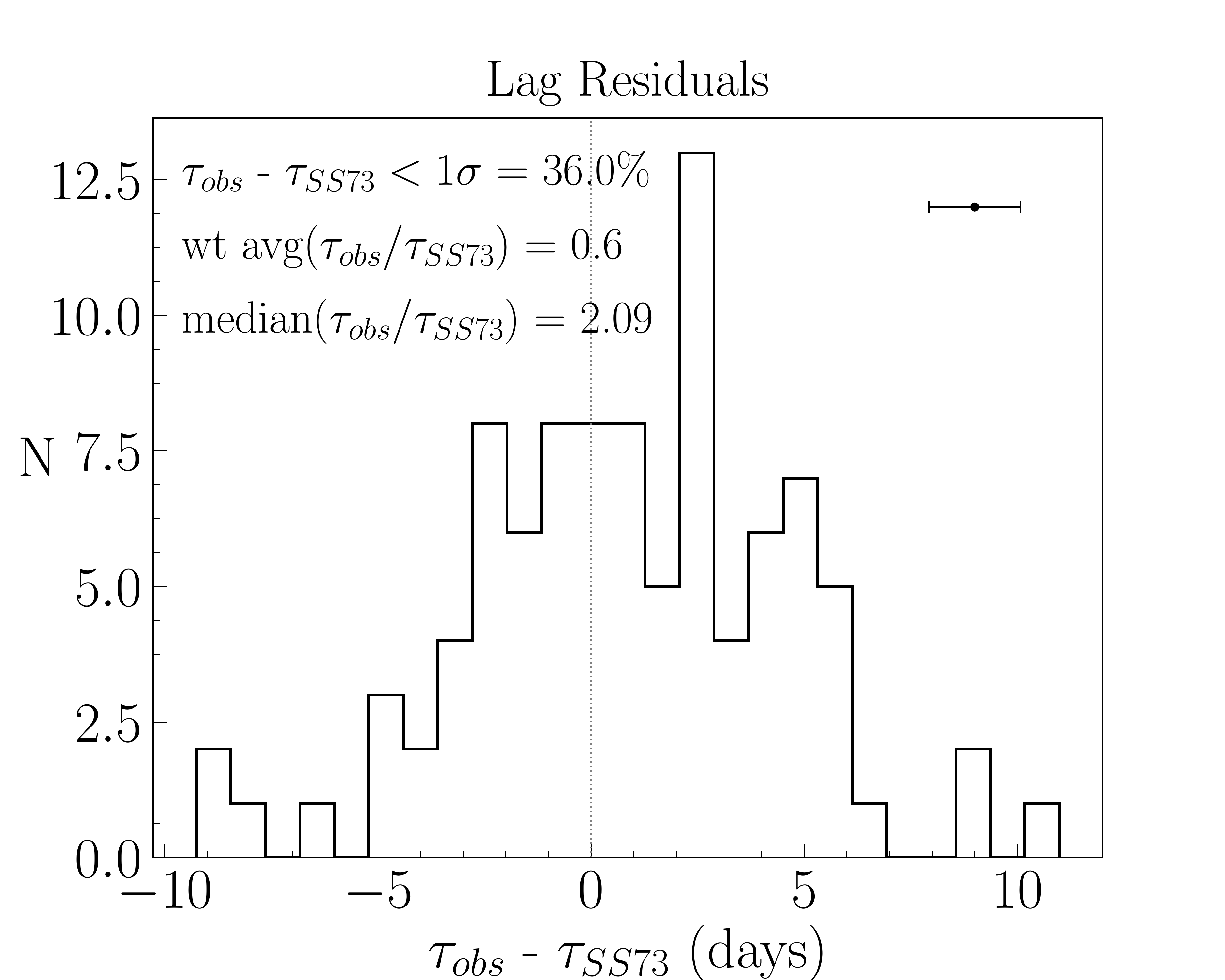}
\caption{Residual of $\tau_{\rm obs}$ and SS73 lags. Here observed lags for the ``well-define" lag sample computed from \texttt{JAVELIN} and model lags are obtained using Equation \eqref{eq:SS73_Fit} based on each object's $M_{BH}$ and $\dot{M}_{\rm BH}$. On average, the observed lags are consistent with the SS73 model lags. But there is considerable scatter, with only 36\% of the observed lags lying within 1$\sigma$ of the model lags.\label{fig:lagcomparison}}
\end{figure}

Figure \ref{fig:lagcomparison} shows a comparison of the observed lags $\tau_{\rm JAV}$ with the analytic model lags $\tau_{\rm SS73}$ calculated from Equation \eqref{eq:SS73_Fit} and \eqref{eq:normalization}. On average, the observed disk sizes are consistent with the SS73 model expectation (including errors on $M_{\rm BH}$ and $\dot{M}$), However, there is large scatter, with only 36\% of the observed ``well-defined" lags lying within 1$\sigma$ of the model lags. The large scatter might indicate that the \texttt{JAVELIN} lag uncertainties are underestimated, or that there are additional important parameters missing from Equations \eqref{eq:SS73_Fit} and \eqref{eq:normalization} such as nonuniform efficiency or orientation. We discuss this issue further in section \ref{sec:Mbh_L3000}.

We perform an initial fit to disk size by first allowing the normalization $\tau_0$ to be the only free parameter and fixing $\beta=4/3$. MCMC then samples the posterior distribution of $\tau_0$. Fitting only the disk normalization based on all of the observed quasar lags in the ``well-defined" sample results in a best-fit disk normalization ${\tau_0}_{\beta = 4/3} = 5.21_{-0.29}^{+0.29}$ days. This is consistent within 1.5~$\sigma$ with the SS73 disk normalization, $\langle\tau_{0}\rangle = 4.78$ days, computed using Equation \eqref{eq:normalization} for the mean $\langle M_{\rm BH}\rangle$ = 8.19 $M_{\odot}$ and $\langle\lambda L_{\lambda 3000}\rangle$ = 44.47 of our sample. We compare our results to those from microlensing \citep{Morgan2010}, and find that our lags are 3-4 times larger than theirs, but this can be attributed to the fact that they use $X = 1$ in Equations \eqref{eq:normalization}, so inflating the SS73 disks of \citet{Morgan2010} by the $X$ = 2.49 will give consistent results with the SS73 expectation (see also \citealt{Tie2018}).  In contrast, \citet{Jiang2017a} find lags that are about 2-3 x larger than SS73. However, the \citet{Jiang2017a} lag sample, by including only significant lags, is biased toward larger lags and thus larger disk sizes. The implication of the bias is less apparent in the recent work by \citep{Mudd2017} where they report consistent lags with SS73. Our measured accretion-disk sizes are similar to those found by \citet{Mudd2017}, in that both our results are broadly consistent with the SS73 model. However our study has the additional advantage of $M_{\rm BH}$ estimates from spectroscopic RM, which we use in Section \ref{sec:Mbh_L3000} to model accretion-disk size as a function of black hole mass and luminosity. In Appendix A, we discuss the effects of observational bias on lag measurements of multi-object quasar samples.  Mixed results are reported for more local quasars e.g., some report lags that are too big \citep{Fausnaugh2016, Fausnaugh2018, Edelson2015, Edelson2017} and some report lags that are close to the SS73 expectation \citet{McHardy2018}. These results may be due to local objects from the NGC-sample are probing the biased tail of the quasar distribution.


\begin{figure}
\includegraphics[scale = 0.28]{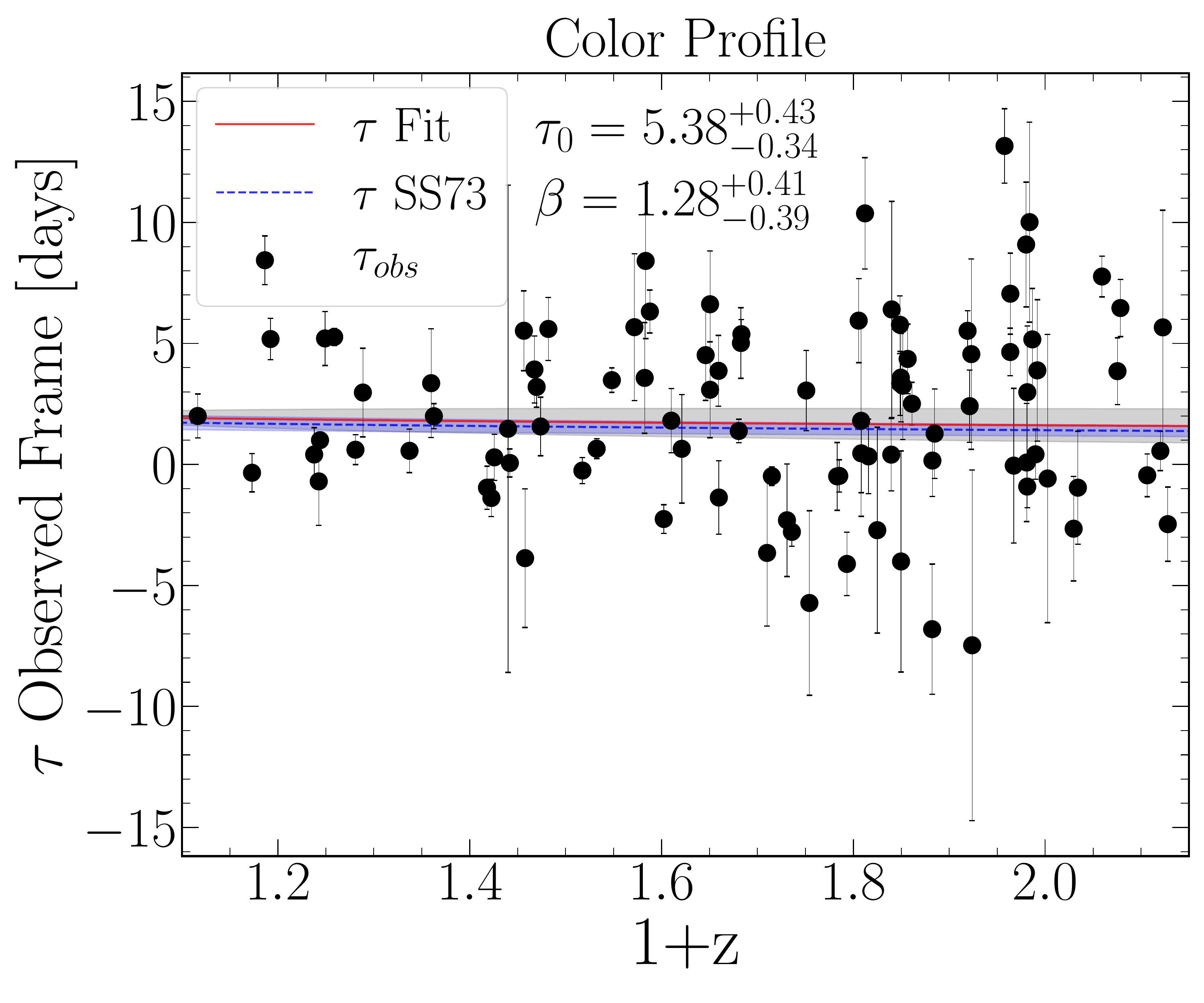}
\caption{Observed lags versus (1+z), fitting a simple accretion disk model with disk normalization $\tau_0$ and wavelength scaling $\beta$ for our sample of ``well-defined" lags. The red line indicates the best-fit disk and the shaded grey region is the propagated error in the best-fit model. The blue line and blue-shaded region shows the SS73 disk model from Equation \eqref{eq:SS73_Fit} and its propagated error with both disk size and wavelength scaling as free parameters. We find a best fit ${\tau_0} = 5.38_{-0.34}^{+0.43}$ and ${\beta} = 1.28_{-0.39}^{+0.41}$ consistent with the SS73 expectation.\label{fig:bowtie_tau0_B}}
\end{figure}
\begin{figure}
\centering
\includegraphics[scale = 0.23]{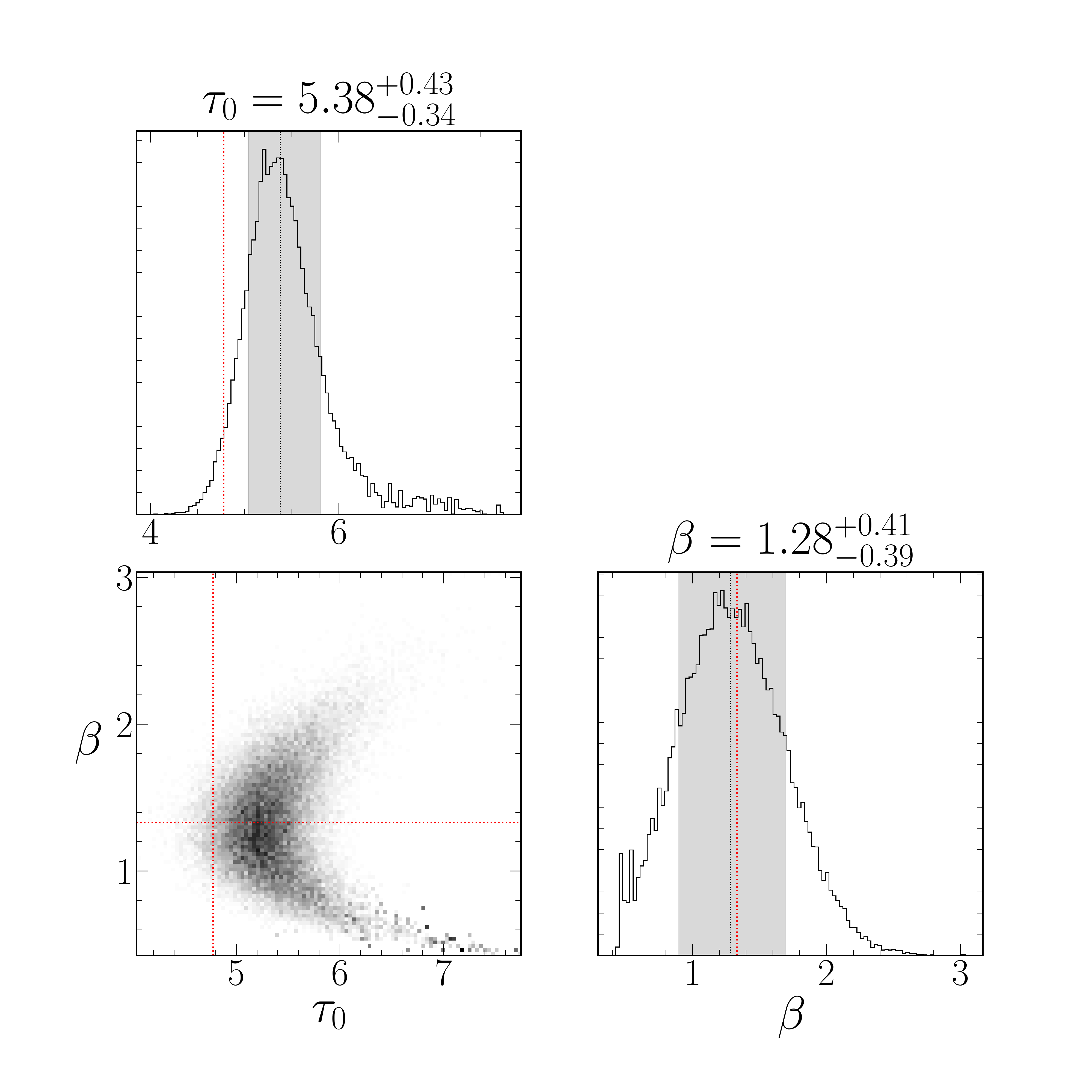}
\caption{Posterior distribution for disk normalization $\tau_0$ and wavelength scaling $\beta$. The shaded gray regions represent the 1$\sigma$ uncertainty of each best-fit parameter and the red dotted line indicates the SS73 expectation using the mean $M_{\rm BH}$ and $\lambda L_{\lambda 3000}$ of our quasar sample.\label{fig:corner_tau0_B}}
\end{figure}
\subsection{Color Profile}\label{sec:beta}
The SS73 accretion disk model predicts a disk structure of $T(R) \propto R^{3/4}$. We measure this temperature profile using wavelength in Equation \eqref{eq:ss73} with a disk size that is characterized by a disk normalization $\tau_0$, wavelength scaling $\beta$, and quasar redshift $z$. In this context, the observed continuum lags are described by:
\begin{equation}\label{eq:ss73}
\tau_{\rm obs} = \tau_0\,(1+z)^{(1 - \beta)}\Big[({\frac{\lambda_i}{9000 \AA}})^{\beta} - ({\frac{\lambda_g}{9000 \AA}})^{\beta}\Big]  
\end{equation}
Although we are only limited to $g$ and $i$ bands in this work, the redshift range of our quasars (0.116 $< z <$ 1.128) provides a broad range of rest-frame wavelengths to test $\beta$, with the best-fit disk size and color profile shown in Figure \ref{fig:bowtie_tau0_B}. The best-fit parameters and errors are determined from the posterior distributions of the MCMC nonlinear regression. We assume the likelihood as a normal distribution, $\mathcal{N}$, centered at observed lags and lag errors as standard deviation. 
\begin{equation}
P(\theta|x) = \mathcal{N}(\tau_{model}|\tau_{obs}, \sigma_{\tau_{obs}})
\end{equation}
Posterior distributions are shown in Figure \ref{fig:corner_tau0_B}: we find $\tau_0 = 5.38^{+0.43}_{-0.34}$ days and $\beta = 1.28^{+0.41}_{-0.39}$.  

Comparing best-fit $\tau_0$ and color profile $\beta$ to the SS73 model indicates that best-fit values are consistent with the SS73 expectation for our sample of mean $M_{\rm BH}$ and $L_{\rm bol}$. Our best-fit color-profile $\beta$ is also consistent within 1$\sigma$ with previous results by \citet{Fausnaugh2016} and \citet{Mudd2017}; further comparison with \citet{Fausnaugh2016} requires multi-band observations as we are only comparing $g$ and $i$ band here. For the remaining portion of this work we will fix $\beta$ to 4/3 in order to focus on the accretion disk connections to $M_{\rm BH}$ and accretion rate.

\vspace{-4mm}
\subsection{Connection to $M_{\rm BH}$ and $\lambda L_{\lambda 3000}$}\label{sec:Mbh_L3000}
Here we examine if our measured continuum lags depend on $M_{\rm BH}^{1/3}$ and $\dot{M}^{1/3}$ as indicated by the SS73 model. 
Our 95 quasars in the ``well-defined" lag sample have reliable $M_{\rm BH}$ estimates using the RM technique for 30 of the quasars and single epoch mass measurements for the remaining 65 quasars: see Table \ref{table1}. To test for connections to $\dot{M}$, we use the observable monochromatic luminosity $\lambda L_{\lambda 3000}$ as a proxy for $\dot{M}$, related as $\dot{M} = L_{bol}/\eta c^2$, with $L_{bol} = 5.15 \lambda L_{\lambda 3000}$. In this context, the observed continuum lags are described by:

\begin{multline}\label{eq:gamma_delta}
\tau_{obs} = \tau_0 \prime \ \left(\frac{M_{\rm BH}} {10^8 M_{\odot}}\right)^{\gamma} \left(\frac{\lambda L_{\lambda 3000}} {10^{44} \, {\rm erg~s}^{-1}}\right)^{\delta}(1+z)^{1-\beta} \\ \Big[\left(\frac{\lambda_i}{9000 \AA}\right)^{\beta}-\left(\frac{\lambda_g}{9000 \AA}\right)^{\beta}\Big]
\end{multline}
\begin{figure*}
\centering
\includegraphics[scale = 0.32]{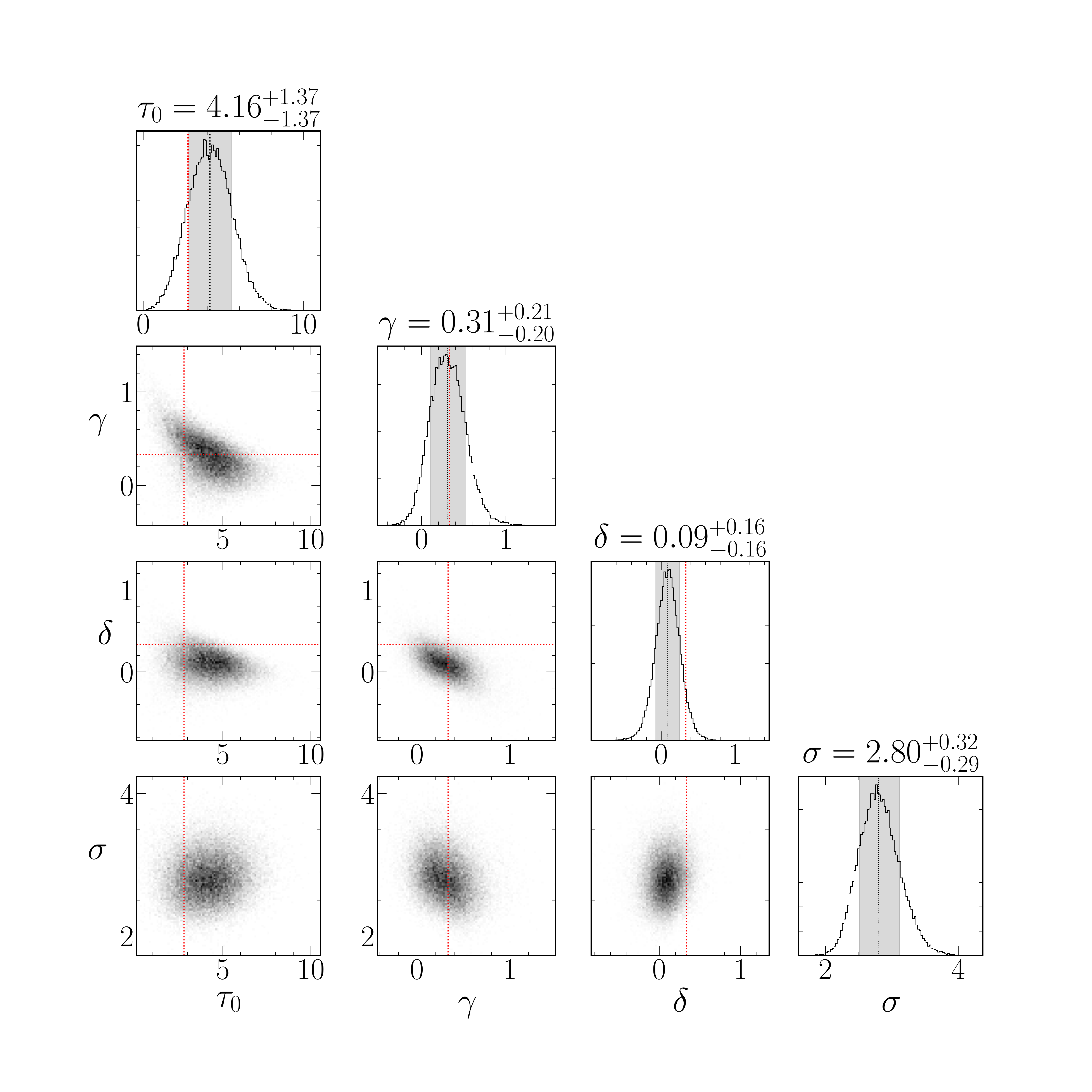}
\caption{Posterior distribution of disk normalization and best-fit $\gamma$ (connection to $M_{\rm BH}$) and $\delta$ (connections to $\lambda L_{\lambda 3000}$) parameter in the disk model presented in Equation~\eqref{eq:gamma_delta} with $\beta = 4/3$ for our sample of ``well-defined" lags.\label{corner_T0GammaDelta}}
\end{figure*}
We perform a new non-linear MCMC regression fit for $\tau_0 \prime$, $\gamma$ and $\delta$. Here $\tau_0 \prime$ has a slightly different form from the previous disk normalization due to different powers in mass and luminosity (i.e., $\tau_0 \prime = \tau_0/M_{\rm BH}^{\gamma} \lambda L_{\lambda 3000}^{\delta}$). We fix $\beta = 4/3$ in Equation \eqref{eq:gamma_delta} and incorporate the measurement uncertainties in $M_{\rm BH}$  reported by \citet{Grier2017}. The uncertainties in RM $M_{\rm BH}$ include a 0.16 dex intrinsic scatter, while for single epoch $M_{\rm BH}$ estimates we assume a 0.4 dex intrinsic scatter \citep{Vestergaard2006,Shen2016a}. We also incorporate the measurement uncertainties while fitting to the observed $\lambda L_{\lambda 3000}$. However the SS73 model predicts disk size as a function of $\dot{M}$ rather than $\lambda L_{\lambda 3000}$, and there is a large scatter between observed luminosity and accretion rate due to uncertainties in bolometric correction and radiative efficiency. This might effectively lead to a larger scatter in the fit, which we measure in the regression fit using an excess dispersion parameter $\sigma$.

\begin{figure*}
\epsscale{1.12}
\plottwo{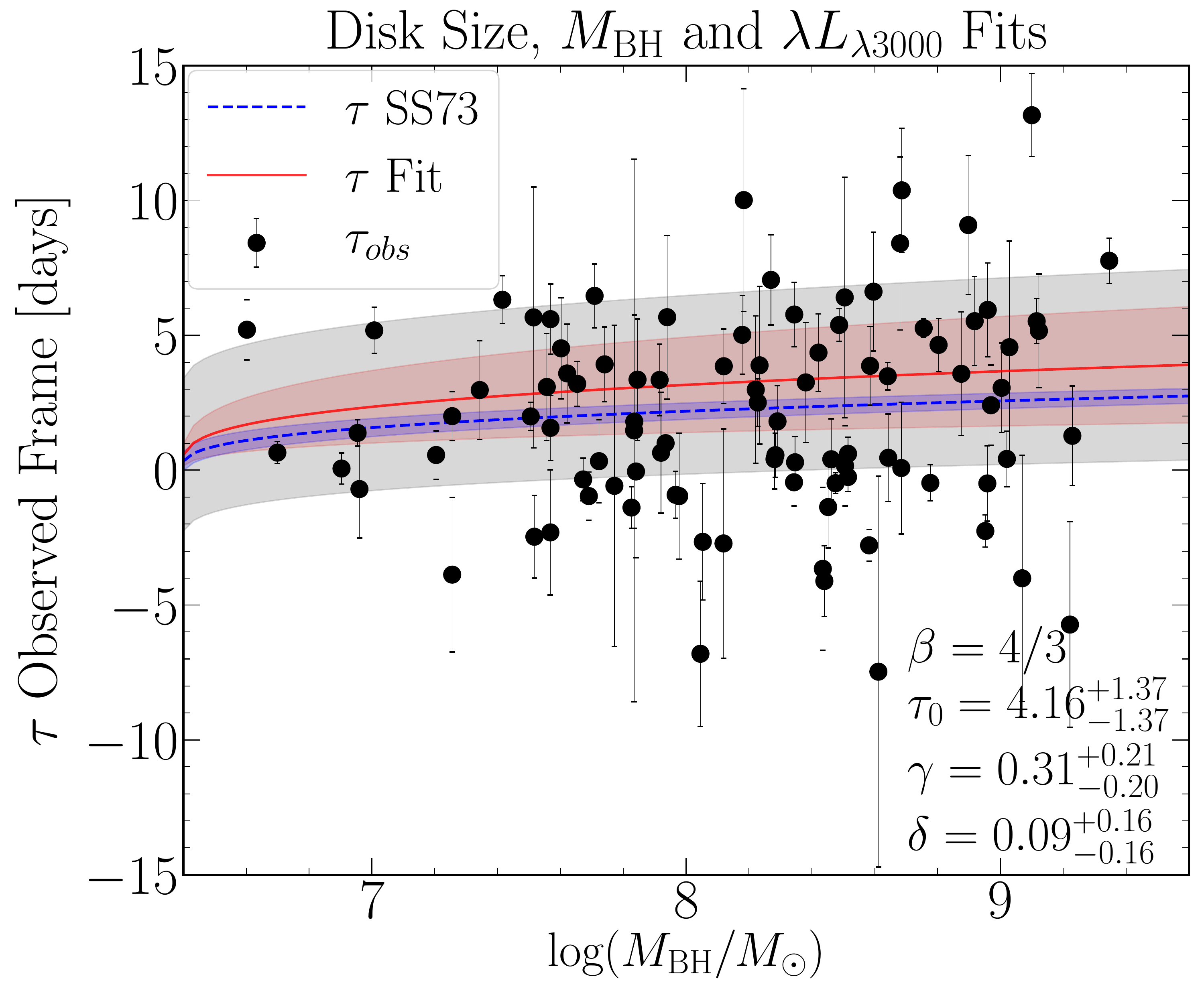}{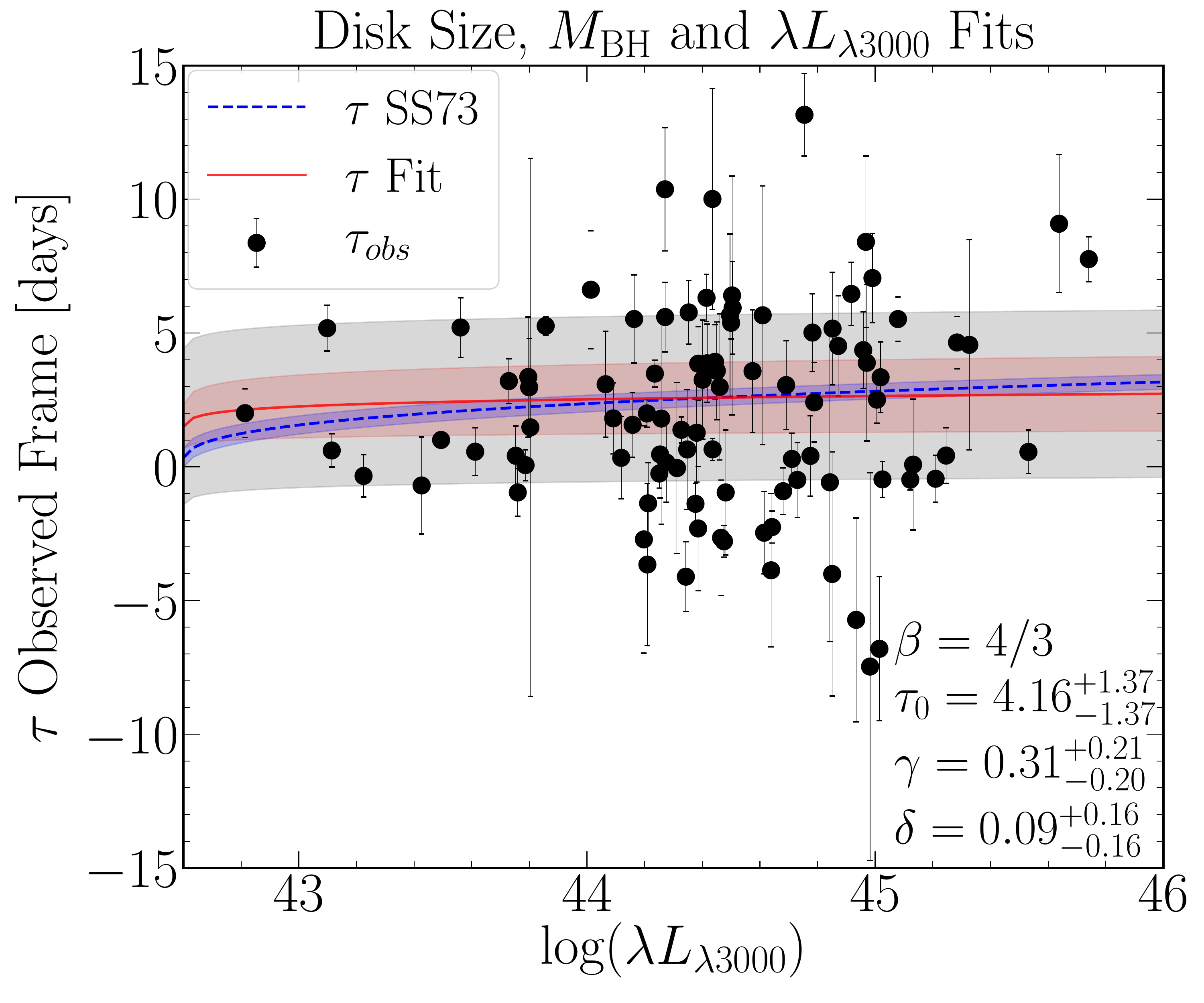}
\caption{\textbf{Left:} Observed ``well-defined" lags versus $M_{\rm BH}$. The best-fit model is shown with solid red line assuming $\tau \propto \tau_{0}\prime \ M^{\gamma} \lambda L_{\lambda 3000}^{\delta}$ for our sample's $M_{\rm BH}$. Here $\tau_{0 SS73}\prime$ is computed from the SS73 theory for our sample's mean redshift and $\lambda L_{\lambda 3000}$. \textbf{Right:} Observed ``well-defined" sample lags versus $\lambda L_{\lambda 3000}$. The best-fit model is shown with solid red line assuming $\tau \propto \tau_{0}\prime \ M^{\gamma} \lambda L_{\lambda 3000}^{\delta}$ for our sample's $\lambda L_{\lambda 3000}$. Here ${\tau\prime}_{0 SS73}$ is computed from the SS73 theory for our sample's mean redshift and $M_{\rm BH}$. In both panels, following our previous consistency-check in \ref{sec:beta} to the SS73, we have assumed $\beta=$ 4/3. The red shading illustrates the propagated error associated with the best-fit parameters, $\tau_0, \gamma, \delta$ and average error in $M_{\rm BH}$ for the plot in the left and average error in $\lambda L_{\lambda 3000}$ for the plot in the right. The gray shading additionally includes the scatter contribution from the excess dispersion $\sigma = 2.8$~days. The blue dashed line illustrates the SS73 disk model as is presented in Equation \eqref{eq:gamma_delta} with $\gamma = \delta$ = 1/3, with blue shading indicating the error contribution from the average $M_{\rm BH}$ uncertainty at left, and the average $\dot{M}_{\rm BH}$ at right (including 0.5 dex scatter for converting from $\lambda L_{\lambda 3000}$ to $\dot{M}_{\rm BH}$).\label{fig:bowtie_Mbh_L3000}}
\end{figure*}

The result of our 3-parameter disk model to the ``well-defined" sample is illustrated in Figures \ref{corner_T0GammaDelta} and \ref{fig:bowtie_Mbh_L3000}.

With disk size parametrized as $\tau_0 \prime$, $M_{\rm BH}^{\gamma}$, and $\lambda L_{\lambda 3000}^{\delta}$ (Equation \ref{eq:gamma_delta}), we find best-fit $\tau_0 \prime = 4.16_{-1.37}^{+1.37}$ days, $\gamma~=~0.31_{-0.20}^{+0.21}$ and $\delta = 0.09_{-0.16}^{+0.16}$. Both $\gamma$ and $\delta$ parameters are poorly constrained, although the mass dependence is $>1\sigma$ different from zero and is fully consistent with the SS73 expectation $\gamma = 1/3$. Our fit indicates that luminosity, $\lambda L_{\lambda 3000}$, on the other hand, is less necessary for the fit, differing from the SS73 expectation by $1.5\sigma$. A more accurate measurement of $\dot{M}_{\rm BH}$ could improve the consistency (i.e., in Equation 1).

Our best-fit parameters include an intrinsic excess dispersion of 2.8 days. This could indicate that the JAVELIN lag errors are underestimated, although the good agreement with \texttt{ICCF} lags in Figure~7 suggests that this is unlikely. Alternatively, individual quasars may have diverse disk emission profiles, with a range of orientation and/or radiative transfer effects that change the $X$ factor in our parameterization \citep{Hall2018}. Some quasars may also have significant continuum emission from a diffuse BLR component, making the measured interband lags differ from pure accretion disk continuum emission \citep{Cackett2018, Edelson2019}.
A non-uniform bolometric correction or radiative efficiency might also lead to scatter in our best-fit disk size as a function of monochromatic luminosity (Equation \eqref{eq:ss73}), although this would have to be as large as 1.8 dex to explain the entirety of the excess scatter measured of $d\tau/\tau= 1.35$ in our regression fit. Finally, it is possible that the SS73 model is a good average description for quasar disks even as individual objects have large variation in their disk structure not captured by the model.

\section{Summary}
We have used continuum RM to study the accretion disks of 222 quasars from the SDSS-RM survey. The selected sample has the advantage of reliable black hole mass measurements from the first year of SDSS-RM monitoring program \citep{Grier2017}. In this work, we used photometric continuum light curves in $g$ and $i$-band to study the accretion disk size and structure of quasars. 

We used \texttt{JAVELIN} to compute lags between $g$ and $i$-band light curves for our 222 quasars. We applied several different significance criteria to obtain a subset of 95 ``well-defined" continuum lags.

Purely comparing our observed lags to those expected from the SS73 model we find a mean deviation of 0.9 days larger than SS73 expectation with 36$\%$ of the ``well-defined" lags consistent within $\pm 1\sigma$ of the SS73 model expectation. 
We perform non-linear MCMC regression to fit our observed lags and compare them to standard SS73 model. Our findings are as follows:
\begin{enumerate}
\item Disk size: Our best-fit disk normalization is consistent with the theoretical value from SS73 within 1.5$~\sigma$. This is in contrast to previous works; possibly due to observational bias (as discussed in Appendix A).
\item Color profile: We find wavelength scaling $\beta=1.28_{-0.39}^{+0.41}$ consistent with the SS73 expectation (i.e., $\beta = 4/3$).
\item Mass and luminosity dependence: We assume disk size $\tau \propto M^{\gamma}\, \lambda L_{\lambda 3000}^{\delta}$ and find best-fit mass dependence $\gamma = 0.31_{-0.20}^{+0.21}$ consistent with expectations from SS73 (i.e., 1/3). The best-fit $\lambda L_{\lambda 3000}$  dependence is $\delta = 0.09_{-0.16}^{+0.16}$, 1.4 $\sigma$ consistent with the SS73 expectation but also $<$1$\sigma$ consistent with no correlation between disk size and luminosity.  Our fits have a large excess dispersion of 2.8 days, indicating a diversity of radiative efficiency, disk emission profiles, and/or disk structure in individual quasars.
\end{enumerate}

Our new measurements represent a large advance over previous work. The 95 SDSS-RM quasars with our new continuum lags and previous broad-line lags \citep{Grier2017} represent a factor of $\sim$ 5 increase over previous samples, and also expands the sample of accretion-disk size and black hole mass measurements by an order of magnitude in redshift, mass, and luminosity. Our measured disk sizes are, on average, consistent with the SS73 analytic thin-disk model. But we also find a large range of smaller and larger disk sizes in excess of the measurement uncertainties. This motivates future work to better measure bolometric luminosity and radiative efficiency (i.e., black hole spin) alongside accretion-disk sizes.

Our work also advances the methodology for accretion-disk size measurements from similar ``industrial-scale" multi-object reverberation projects beyond SDSS-RM.  In particular, we advocate a Bayesian approach to the full sample of ``well-defined" lag measurements, rather than restricting analysis to a set of high-SNR lags that are biased by limitations in survey cadence. SDSS-RM is planned to continue in the 2020s with a factor of 5 increase in survey area as part of the SDSS-V Black Hole Mapper project \citep{Kollmeier2018, Ivezic2018}. The Large Synoptic Survey Telescope (LSST) will usher in an entirely new era of time-domain quasar studies, making continuum reverberation mapping possible for thousands of quasars in its deep drilling fields.

\acknowledgments
We thank Michael Fausnaugh for helpful discussion that improved the manuscript. YH and JRT acknowledge support from NASA grant HST-GO-15260. CJG, WNB, and DPS acknowledge support from NSF grant AST-1517113. YS, DAS and JL acknowledge support from an Alfred P. Sloan Research Fellowship (YS) and NSF
grant AST-1715579. KH acknowledges support from STFC grant ST/R000824/1. PH acknowledges support from the Natural Sciences and Engineering Research Council of Canada (NSERC), funding reference number 2017-05983, and from the National Research Council Canada during his sabbatical at NRC Herzberg Astronomy \& Astrophysics. LCH was supported by the National Key R\&D Program of China (2016YFA0400702) and the National Science Foundation of China (11473002, 11721303).

Funding for SDSS-III was provided by the Alfred P. Sloan Foundation, the Participating Institutions, the National Science Foundation, and the U.S. Department of Energy Office of Science. The SDSS-III web site is http://www.sdss3.org/.  SDSS-III was managed by the Astrophysical Research Consortium for the Participating Institutions of the SDSS-III Collaboration including the University of Arizona, the Brazilian Participation Group, Brookhaven National Laboratory, Carnegie Mellon University, University of Florida, the French Participation Group, the German Participation Group, Harvard University, the Instituto de Astrofisica de Canarias, the Michigan State/Notre Dame/JINA Participation Group, Johns Hopkins University, Lawrence Berkeley National Laboratory, Max Planck Institute for Astrophysics, Max Planck Institute for Extraterrestrial Physics, New Mexico State University, New York University, Ohio State University, Pennsylvania State University, University of Portsmouth, Princeton University, the Spanish Participation Group, University of Tokyo, University of Utah, Vanderbilt University, University of Virginia, University of Washington, and Yale University.

We thank the Bok and CFHT Canadian, Chinese, and French TACs for their support. This research uses data obtained through the Telescope Access Program (TAP), which is funded by the National Astronomical Observatories, Chinese Academy of Sciences, and the Special Fund for Astronomy from the Ministry of Finance in China.  This work is based on observations obtained with MegaPrime/MegaCam, a joint project of CFHT and CEA/DAPNIA, at the Canada-France-Hawaii Telescope (CFHT) which is operated by the National Research Council (NRC) of Canada, the Institut National des Sciences de l’Univers of the Centre National de la Recherche Scientifique of France, and the University of Hawaii. The authors wish to recognize and acknowledge the very significant cultural role and reverence that the summit of Maunakea has always had within the indigenous Hawaiian community. The astronomical community is most fortunate to have the opportunity to conduct observations from this mountain.

\appendix
\section{Selection Bias}
\begin{figure*}
\epsscale{1.1}
\plottwo{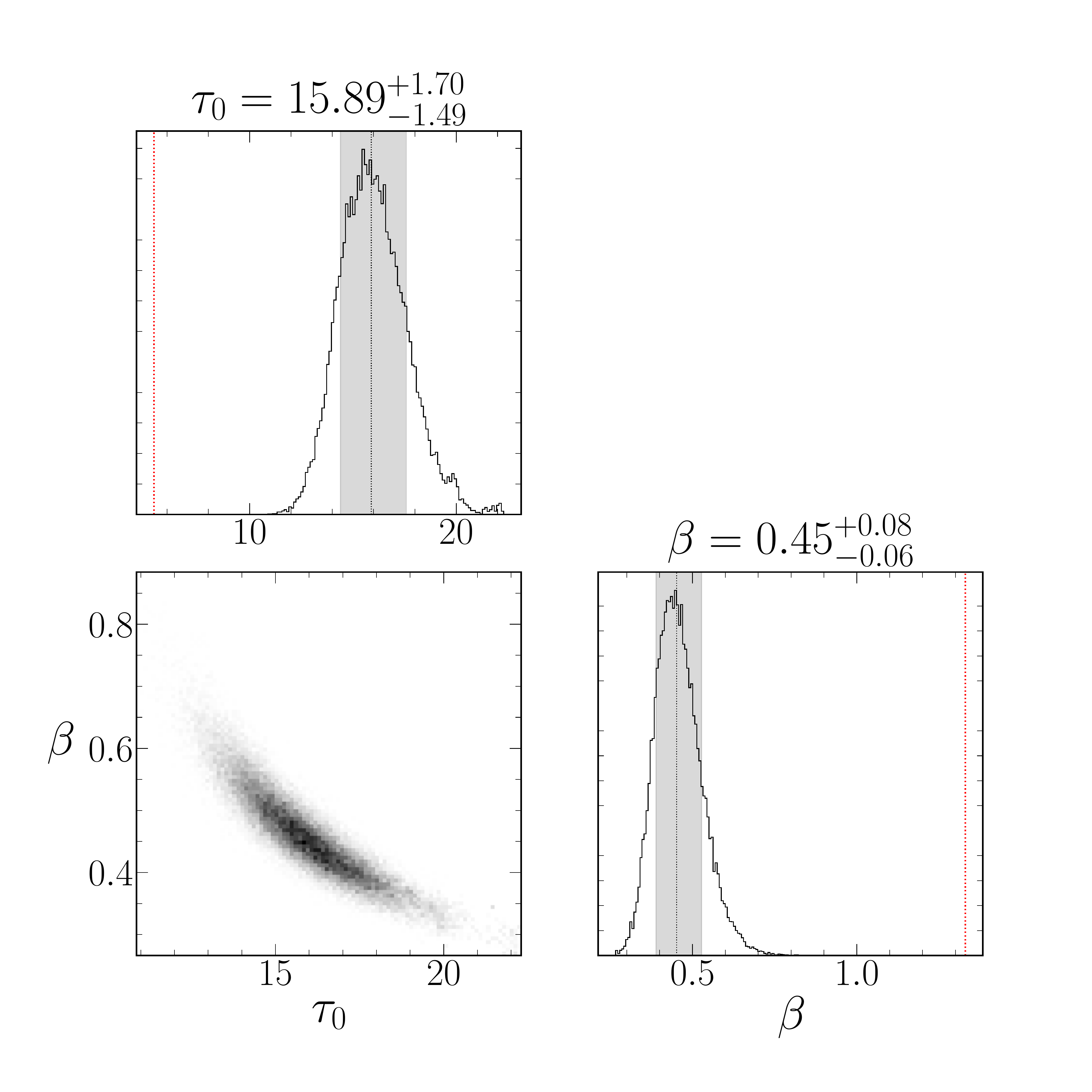}{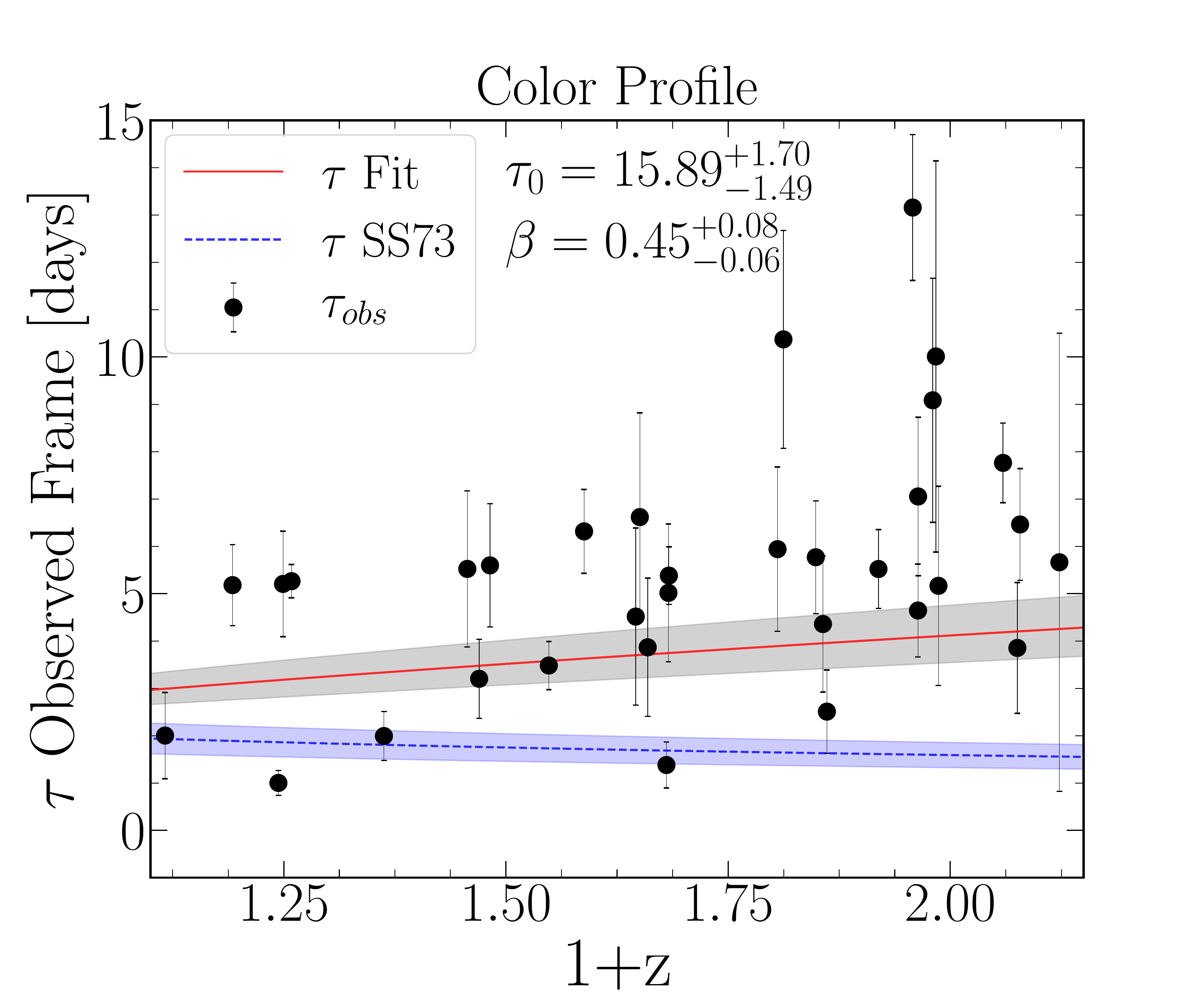}
\caption{\textbf{Left:} Prior distribution for disk normalization $\tau_0$ and wavelength scaling $\beta$ using only the\textit{``high-SNR" lag sample}. The shaded gray region shows the 1$\sigma$ uncertainty of each best-fit parameter and the red dotted line indicates the SS73 expectation using the mean $M_{\rm BH}$ and $\lambda L_{\lambda 3000}$ of our ``high-SNR" sample. \textbf{Right:} Observed ``high-SNR"\texttt{JAVELIN} lags versus (1+z). Best-fit model using $\beta$ and $\tau_0$ is shown with a solid red line and the shading illustrates the 1$\sigma$ propagated errors on $\tau_{model}$ from the MCMC parameter errors. The blue dashed line shows the SS73 model from Equation \ref{eq:SS73_Fit}.\label{fig:strong_tau0_B}}
\end{figure*}

\begin{figure*}
\epsscale{1.1}
\plottwo{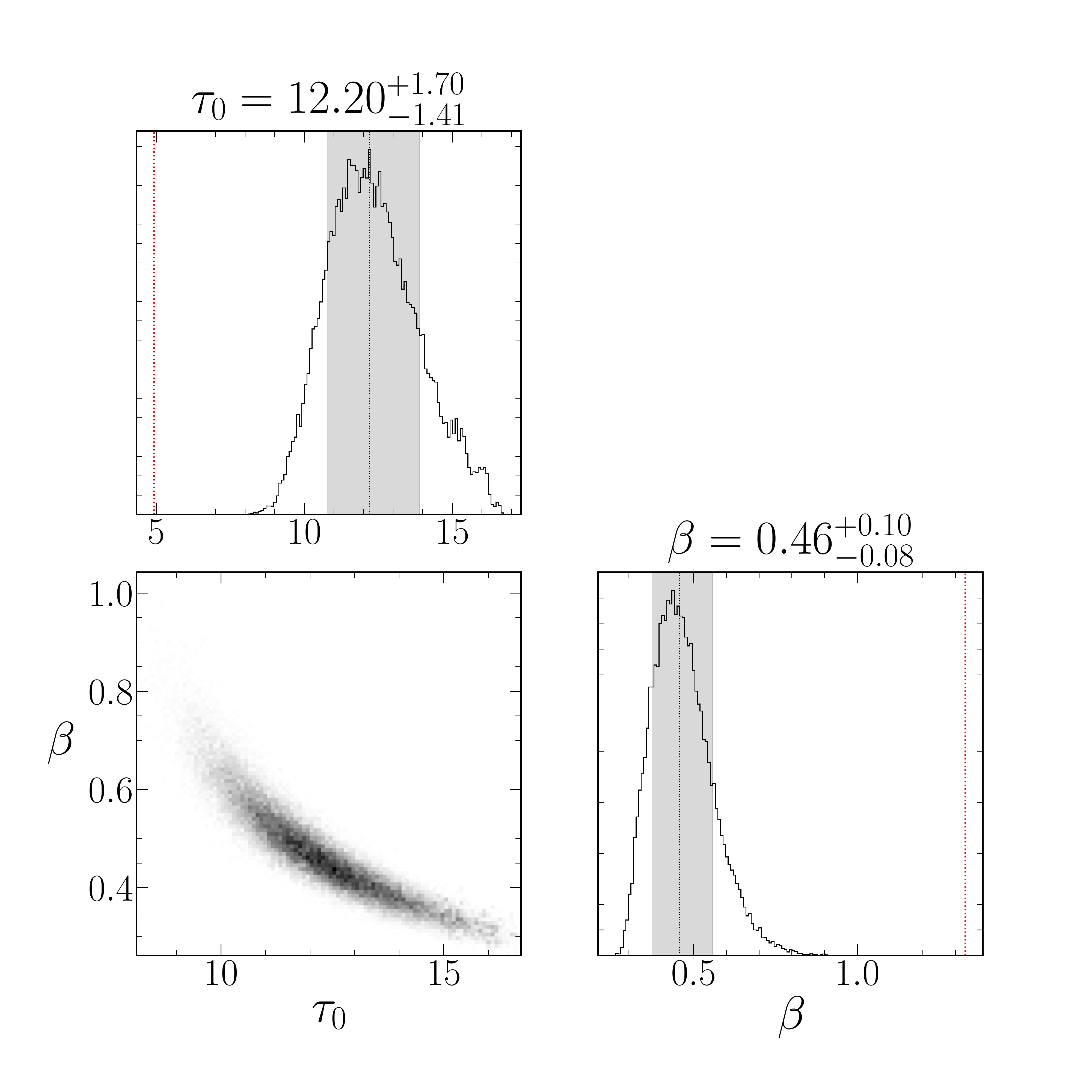}{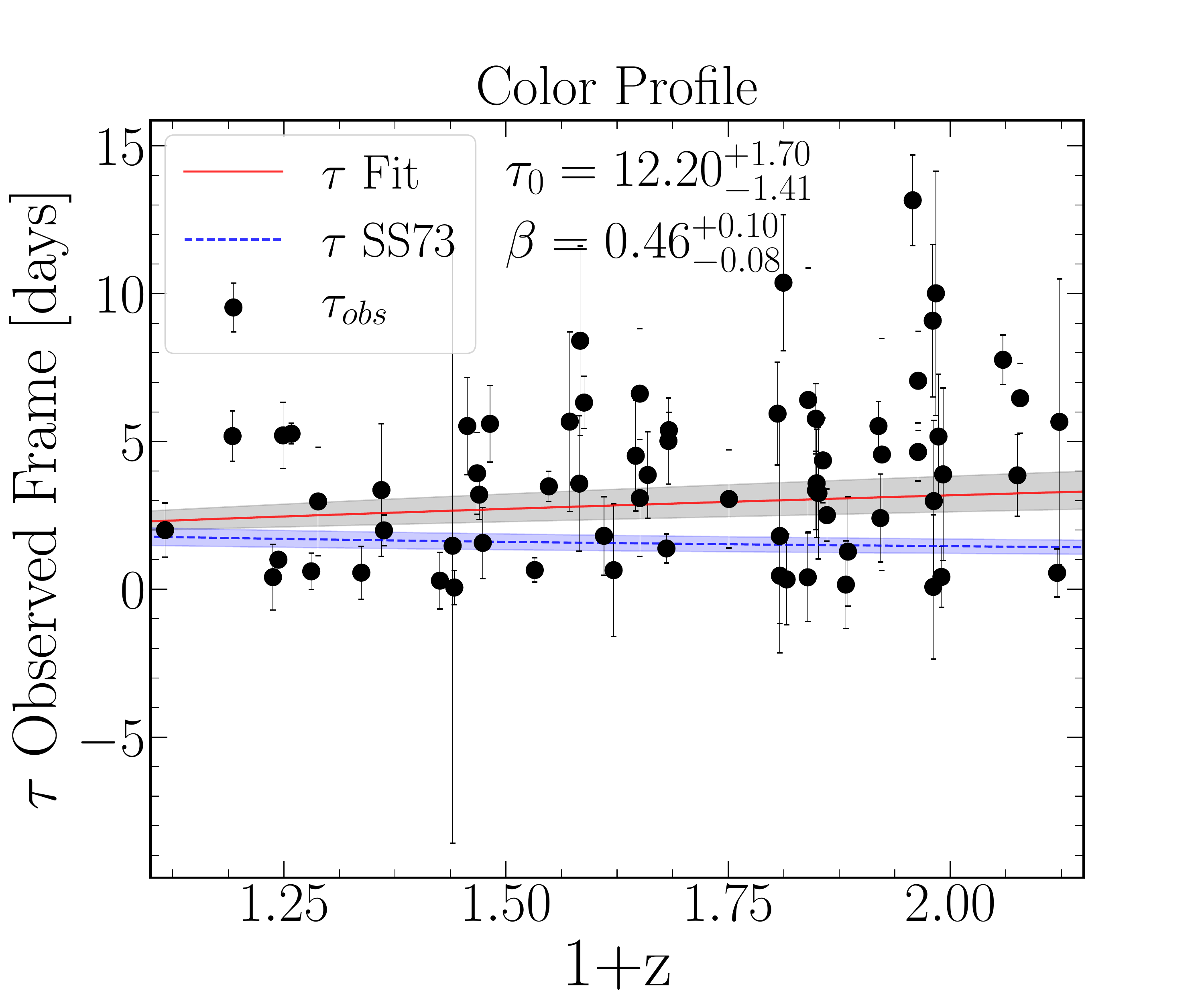}
\caption{\textbf{Left:} Prior distribution for disk normalization $\tau_0$ and wavelength scaling $\beta$ using only the \textit{positive``well-defined" lag sample}. The shaded gray region shows the 1$\sigma$ uncertainty of each best-fit parameter and the red dotted line indicates the SS73 expectation using the mean $M_{\rm BH}$ and $\lambda L_{\lambda 3000}$ of our positive, ``high-SNR" sample. \textbf{Right:} Observed positive, ``well-defined" \texttt{JAVELIN} lags versus (1+z). The best-fit model using only $\beta$ and $\tau_0$ is shown with a solid red line and the shading illustrates the 1$\sigma$ propagated errors on $\tau_{model}$ from the MCMC parameter errors. The blue dashed line shows the SS73 model from Equation \eqref{eq:SS73_Fit}.\label{fig:positive_tau0_B}}
\end{figure*}

We take a Bayesian approach in Section \ref{sec:discussion} and fit all 95 quasars with ``well-defined" lags (see Section \ref{criterion}), including those that are consistent with zero lag. However, if we instead fit only the high-SNR lags (``well-defined" and lag SNR $> 2\sigma$) we find disks that are $\sim$~3.2 times larger than SS73 and a nearly-flat color profile $\beta = 0.4$ shown in Figure \ref{fig:strong_tau0_B}. Additionally, we test for ``well-defined" and positive lags and find disks that are $\sim$2.5 times larger than expectation by the SS73, see Figure \ref{fig:positive_tau0_B}.

The high-SNR sample is biased to large lags, as the SDSS-RM cadence (averaging 4 days) sets a minimum detectable lag. This biases the disk fits to large values. Similar bias is likely to affect the main sample in \citep{Jiang2017a} as they used only positive lags in their fits. We reproduce the same qualitative effects if we limit our sample to only positive lags, see Figure \ref{fig:positive_tau0_B}.

Our larger ``well-defined" lag sample, on the other hand, is not biased to large lags.  Although the sample includes many lags that are formally consistent with zero, the lags are more likely to be positive than negative, as shown in Figure \ref{fig:javdiagnostic}. This indicates that the lags are likely the result of genuine reverberation but are just smaller than detectable by the SDSS-RM cadence (average of 4 days).  In other words, the ``well-defined" sample includes many lags that have poor SNR but are constrained to be small. It is important to include such lags in the accretion-disk fits to avoid a bias to large disk sizes.

\section{Increased short-timescale variability}
The SDSS-RM time monitoring observations (see Section \ref{sec:data}) are limited by somewhat sparsely sampled data with a median cadence of 4 days. Although the expected quasar variability on such short time scales is relatively low \citep{Mushotzky2011,MacLeod2012}, our measured lags are fundamentally limited by the observational cadence. Here we validate that quasar fluctuations on timesales shorter than our observation cadence do not affect our lag measurements by constructing synthetic light curve that have extreme variability in between each measured points. 

To construct our new synthetic light curves, we take each consecutive measured flux pair $f(t_i), f(t_{i+1})$ and randomly select an inter-point expectation flux $f(t_{i+1/2})$, where $t_{i+1/2}= t_i + (t_{i+1}-t_i)/2$, from the \texttt{JAVELIN} DRW model normal distribution. We then increase (or decrease) each randomly-selected flux to a new flux $f(t_{i+1/2}) + \delta f$, where $\delta f \equiv 1/2 (f(t_{i+1}) - f(t_i))$, i.e., varying by half the difference between consecutive pairs of measured fluxes. This is equivalent to a short-timescale variability PSD of $\alpha=-1$: an extreme variability case compared to a DRW ($\alpha=-2$) and to the low measured short-timescale variability of $\alpha \simeq -3$ \citep{Mushotzky2011}.  We also perturbed the new inter-point flux by the average flux uncertainty of the measured surrounding flux pair. The final synthetic light curve is then the combination of both the measured light curve and the new inter-point fluxes.

\begin{figure*}
\centering
\includegraphics[scale = 0.30]{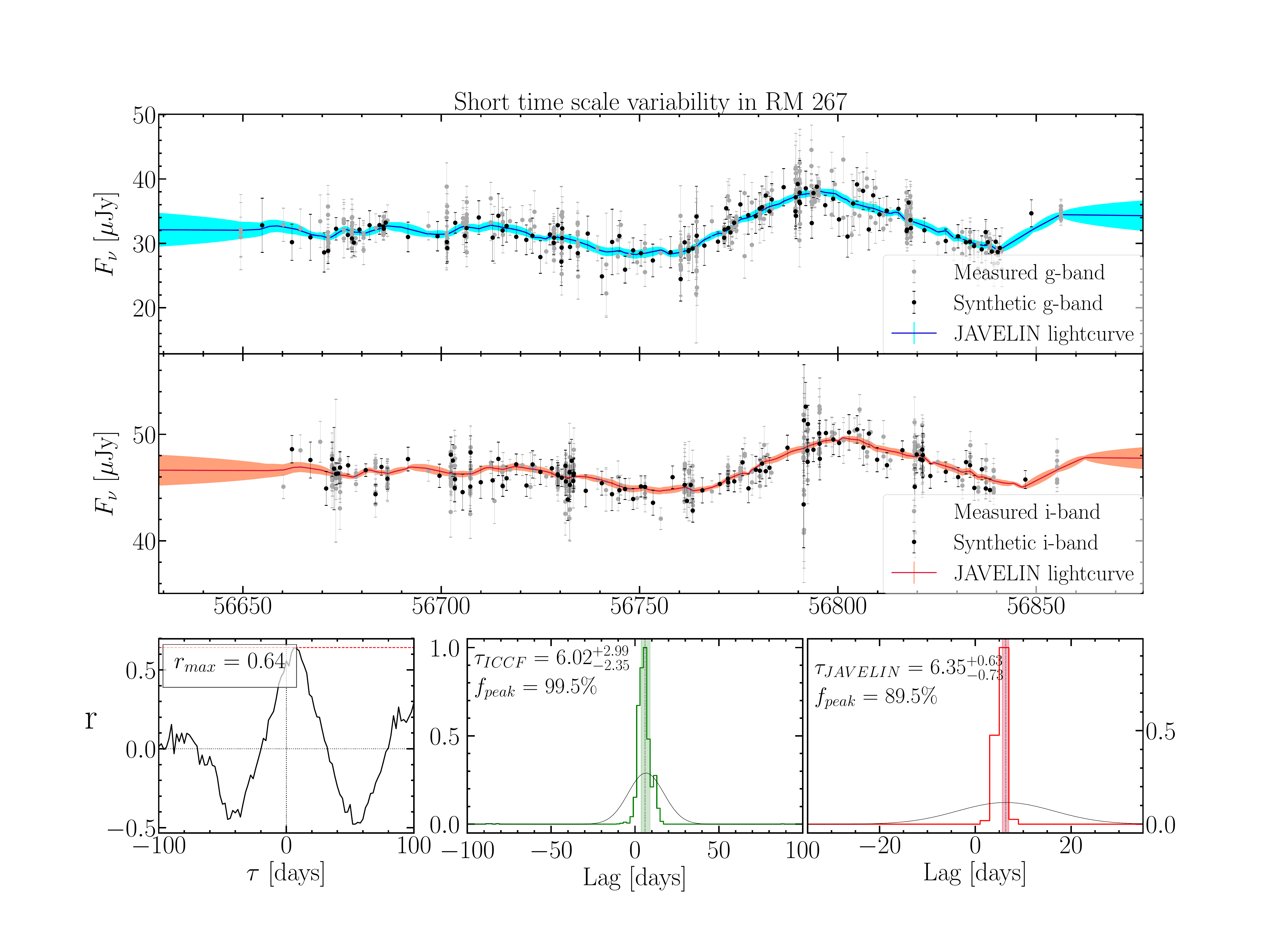}
\caption{\textbf{Top:}  Synthetic Continuum $g$ (blue) and $i$-band (red) light curves with increased short timescale variability in between observations. The inter-point variability was increased between each consecutive pair of observations. The inter-point flux was randomly selected between consecutive observations using the \texttt{Javelin} DRW model normal distribution increased (decreased) by half the difference of observation pair. \textbf{Bottom:} We find the same lag with both \texttt{Javelin} and \texttt{ICCF} methods.\label{fig:synth}}
\end{figure*}

We build synthetic light curves for all of our targets in the ``well-defined" sample and use \texttt{JAVELIN} to measure lags as described in Section \ref{lag_ID}. We find that the synthetic light curves have lags that are statistically consistent with the original lags measured from the observed lightcurves. Figure \ref{fig:synth} illustrates the synthetic light curve with extreme short-timescale variability" for RM 267 (the same target as Figure 4) and the measured \texttt{JAVELIN} and \texttt{ICCF} lag probability distributions.

\end{document}